\documentclass[aps, prb, reprint, superscriptaddress]{revtex4-1}

\usepackage{todonotes}
\usepackage{bm}
\usepackage{graphicx}
\usepackage{amsmath}
\usepackage{mathbbol}
\usepackage{amssymb}
\usepackage[normalem]{ulem}
\usepackage{color}
\usepackage{pdfpages}

\makeatletter
\AtBeginDocument{\let\LS@rot\@undefined}
\makeatother

\begin{document}

\title{Dynamics of skyrmionic states in confined helimagnetic nanostructures}

\author{Marijan Beg}
\email{m.beg@soton.ac.uk}
\author{Maximilian Albert}
\author{Marc-Antonio Bisotti}
\author{David Cort\'{e}s-Ortu\~{n}o}
\affiliation{Faculty of Engineering and the Environment, University of Southampton, Southampton SO17 1BJ, United Kingdom}
\author{Weiwei Wang}
\affiliation{Faculty of Engineering and the Environment, University of Southampton, Southampton SO17 1BJ, United Kingdom}
\affiliation{Department of Physics, Ningbo University, Ningbo, 315211, China}
\author{Rebecca Carey}
\author{Mark Vousden}
\author{Ondrej Hovorka}
\affiliation{Faculty of Engineering and the Environment, University of Southampton, Southampton SO17 1BJ, United Kingdom}
\author{Chiara Ciccarelli}
\affiliation{Cavendish Laboratory, University of Cambridge, Cambridge CB3 0HE, United Kingdom}
\author{Charles S. Spencer}
\author{Christopher H. Marrows}
\affiliation{School of Physics and Astronomy, University of Leeds, Leeds LS2 9JT, United Kingdom}
\author{Hans Fangohr}
\email{h.fangohr@soton.ac.uk}
\affiliation{Faculty of Engineering and the Environment, University of Southampton, Southampton SO17 1BJ, United Kingdom}

\begin{abstract}
In confined helimagnetic nanostructures, skyrmionic states in the form of incomplete and isolated skyrmion states can emerge as the ground state in absence of both external magnetic field and magnetocrystalline anisotropy. In this work, we study the dynamic properties (resonance frequencies and corresponding eigenmodes) of skyrmionic states in thin film FeGe disk samples. We employ two different methods in finite-element based micromagnetic simulation: eigenvalue and ringdown method. The eigenvalue method allows us to identify all resonance frequencies and corresponding eigenmodes that can exist in the simulated system. However, using a particular experimentally feasible excitation can excite only a limited set of eigenmodes. Because of that, we perform ringdown simulations that resemble the experimental setup using both in-plane and out-of-plane excitations. In addition, we report the nonlinear dependence of resonance frequencies on the external magnetic bias field and disk sample diameter and discuss the possible reversal mode of skyrmionic states. We compare the power spectral densities of incomplete skyrmion and isolated skyrmion states and observe several key differences that can contribute to the experimental identification of the state present in the sample. We measure the FeGe Gilbert damping, and using its value we determine what eigenmodes can be expected to be observed in experiments. Finally, we show that neglecting the demagnetisation energy contribution or ignoring the magnetisation variation in the out-of-film direction -- although not changing the eigenmode's magnetisation dynamics significantly -- changes their resonance frequencies substantially. Apart from contributing to the understanding of skyrmionic states physics, this systematic work can be used as a guide for the experimental identification of skyrmionic states in confined helimagnetic nanostructures.
\end{abstract}

\maketitle
\section{Introduction}
Dzyaloshinskii-Moriya interactions~\cite{Dzyaloshinsky1958, Moriya1960} (DMI) may occur in magnetic systems that lack some type of inversion symmetry. The inversion asymmetry can be present in the magnetic system either because of a non-centrosymmetric crystal lattice~\cite{Moriya1960} (helimagnetic material) or due to the interfaces between different materials which inherently lack inversion symmetry.~\cite{Fert1980, Crepieux1998} Consequently, the DMI can be classified either as bulk or interfacial. The DMI favours magnetic moments at neighbouring lattice sites to be perpendicular to each other (in plane that is perpendicular to the Dzyaloshinskii vector), which is in contrast to the symmetric ferromagnetic exchange interaction which tends to align them parallel. When acting together, these two interactions mutually compete and find a compromise in the twist between two neighbouring magnetic moments, which allows a rich variety of different magnetisation textures. One of them is a skyrmion configuration with very promising properties~\cite{Heinze2011, Kanazawa2012, Romming2013, Jonietz2010, Yu2012} for the development of future high-density, power-efficient storage~\cite{Kiselev2011, Fert2013} and logic~\cite{Zhang2015} devices.

After it was predicted~\cite{Bogdanov1989, Bogdanov1999, Rossler2006} that magnetic skyrmions can emerge in the presence of DMI, skyrmions were observed in magnetic systems with both bulk~\cite{Muhlbauer2009, Yu2011, Yu2010, Seki2012, Kanazawa2012} and interfacial~\cite{Heinze2011, Romming2013, Sonntag2014} types of DMI. However, all studies of helimagnetic (bulk DMI) materials required an external magnetic field to be applied in order to stabilise skyrmions. Recently, a systematic micromagnetic study~\cite{Beg2015} reported all equilibrium states that can emerge in confined helimagnetic nanostructures and identified the lowest energy (ground) states. In particular, this study reported that in confined helimagnetic nanostructures two different skyrmionic states can emerge as the ground state in absence of both external magnetic field and magnetocrystalline anisotropy. One state does not contain a complete spin rotation, whereas the other state contains one full spin rotation along the disk sample diameter, plus an additional magnetisation tilting at the boundary due to the specific boundary conditions.~\cite{Rohart2013} We refer to these configurations as incomplete skyrmion (iSk) and isolated skyrmion (Sk) states, respectively.~\cite{Beg2015} In addition, the same study showed that the higher-order target (T) state with two complete spin rotations along the disk sample diameter can emerge as a metastable state at zero external magnetic field.

Understanding the dynamic response of skyrmionic states in confined helimagnetic nanostructures is of importance both from the aspect of fundamental physics as well as for their manipulation. In this work, we explore the dynamics of all three equilibrium skyrmionic states using a full three-dimensional model which includes the demagnetisation energy contribution and does not assume the translational invariance of magnetisation in the out-of-film direction. A similar dynamics simulation study was performed for the isolated skyrmion breathing eigenmodes~\cite{Kim2014} in confined two-dimensional samples with interfacial DMI; and high-frequency skyrmion spin excitations were analytically studied in thin cylindrical dots.~\cite{Gareeva2016} The low-frequency (two lateral and one breathing) eigenmodes were reported in two-dimensional simulations of a hexagonal skyrmion lattice,~\cite{Mochizuki2012} where the demagnetisation energy was neglected. Later, microwave absorption measurements explored the low frequency eigenmodes in Cu$_{2}$OSeO$_{3}$,~\cite{Onose2012, Okamura2013, Schwarze2015} Fe$_{1-x}$Co$_{x}$Si,~\cite{Schwarze2015} and MnSi~\cite{Schwarze2015} helimagnetic bulk samples. In the case of a magnetic bubble~\cite{Buda2001, Moutafis2006, Moutafis2007, Guslienko2015} (magnetisation state stabilised due to the strong uniaxial anisotropy in the absence of DMI) analytic,~\cite{Makhfudz2012} simulation,~\cite{Moutafis2009} and experimental~\cite{Buttner2015} studies reported the existence of two low frequency gyrotropic eigenmodes, suggesting that the skyrmion possesses mass. In contrast, a recent analytic work~\cite{Guslienko2016} suggests that only one gyrotropic eigenmode exists in the confined DMI-induced skyrmion state, whereas another low-frequency lateral eigenmode is interpreted as an azimuthal spin-wave mode~\cite{Guslienko2016}.

Using our full three-dimensional model, employing the eigenvalue~\cite{DAquino2009} method, we compute all existing (both lateral and breathing) eigenmodes below $50 \,\text{GHz}$ in three different skyrmionic states. In addition, using the ringdown~\cite{McMichael2005} method, we determine which eigenmodes can be excited using two different experimentally feasible excitations (in-plane and out-of-plane). In contrast to the magnetic bubble, in the confined DMI stabilised skyrmionic states we find the existence of only one low frequency gyrotropic eigenmode. We also demonstrate the nonlinear dependence of eigenmode frequencies on the external magnetic bias field and the disk sample diameter, and show that the gyrotropic eigenmode might be the reversal mode of skyrmionic states. After we identify all eigenmodes of incomplete skyrmion (iSk) and isolated skyrmion (Sk) ground states, we compare their power spectral densities (PSDs) in the same sample at different external magnetic field values. We discuss the comparisons and observe several key differences that can contribute to the experimental identification of the state present in the studied sample. Although we base this study on the specific helimagnetic material FeGe, in order to make this study relevant to any helimagnetic material, we need to determine as many resonant frequencies as possible that can be detected using a specific excitation. Because of that, we need to reduce the linewidth and allow sufficient separation between peaks in the power spectral density (computed using the ringdown method). Consequently, in the first part of this work, we use the Gilbert damping~\cite{Kim2014} $\alpha' = 0.002$. After we identify all resonance frequencies and corresponding eigenmodes using $\alpha'$, we experimentally measure the real value of FeGe Gilbert damping and use it to determine which (out of all previously identified eigenmodes) can be experimentally detected in the FeGe sample. Finally, we investigate how the demagnetisation energy contribution and magnetisation variation in the out-of-film direction affect the dynamics of skyrmionic states. We report that although the eigenmode magnetisation dynamics is not significantly affected, the resonance frequencies change substantially, which indicates that ignoring the demagnetisation energy or modelling the thin film helimagnetic samples using two-dimensional meshes is not always justified.

\section{Methods}
\begin{figure}
  \includegraphics{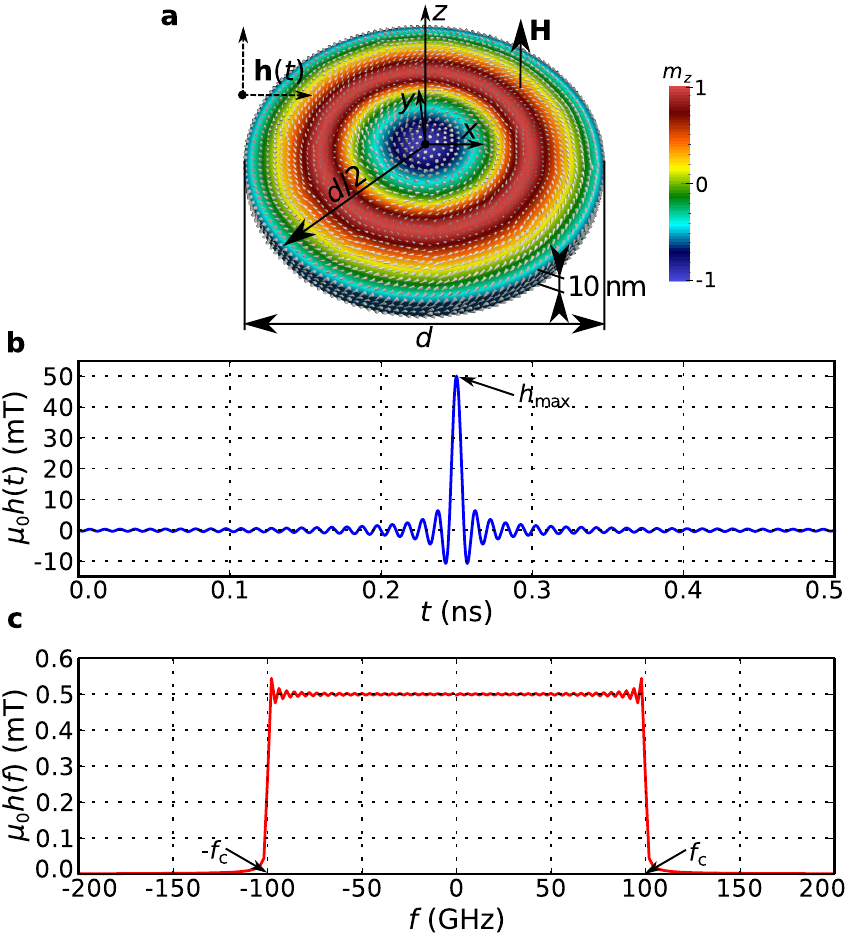}
  \caption{\label{fig:methods} (a)~A thin film FeGe disk sample with $10 \,\text{nm}$ thickness and diameter $d$. An external magnetic bias field $\mathbf{H}$ is applied uniformly and perpendicular to the sample (in the positive $z$~direction). (b)~A cardinal sine wave excitation magnetic field $\mathbf{h}(t)$, used in the ringdown method, is applied for $0.5 \,\text{ns}$ in either in-plane ($\hat{\mathbf{x}}$) or out-of-plane ($\hat{\mathbf{z}}$) direction. (c)~The Fourier transform of excitation field $h(t)$ shows that all eigenmodes (allowed by the used excitation direction) with frequencies lower than $f_\text{c} = 100 \,\text{GHz}$ are excited approximately equally.}
\end{figure}

We simulate a thin film helimagnetic cubic B20 FeGe disk with $10 \,\text{nm}$ thickness and diameter $d$, as shown in Fig.~\ref{fig:methods}(a). The thin film sample is in the $xy$ plane and perpendicular to the $z$~axis. An external magnetic bias field $\mathbf{H}$ is applied uniformly and perpendicular to the sample (in the positive $z$~direction).

The total energy of the system we simulate contains several energy contributions and can be written as
\begin{equation}
  \label{eq:total_energy}
  E = \int \left[ w_\text{ex} + w_\text{dmi} + w_\text{z} + w_\text{d} + w_\text{a} \right] \,\text{d}^{3}r.
\end{equation}
The first term $w_\text{ex} = A \left[ (\nabla m_{x})^{2} + (\nabla m_{y})^{2} + (\nabla m_{z})^{2} \right]$ is the symmetric exchange energy density with material parameter $A$. The unit vector field $\mathbf{m} = \mathbf{m}(\mathbf{r}, t)$, with Cartesian components $m_{x}$, $m_{y}$, and $m_{z}$, represents the magnetisation field $\mathbf{M}(\mathbf{r}, t) = M_\text{s}\mathbf{m}(\mathbf{r}, t)$, where $M_\text{s}$ is the saturation magnetisation. The second term $w_\text{dmi} = D \mathbf{m} \cdot \left(\nabla \times \mathbf{m} \right)$ is the Dzyaloshinskii-Moriya energy density with material parameter $D$, which is obtained by including Lifshitz invariants suitable for materials of the crystallographic class T, such as the cubic B20 FeGe (P$2_{1}3$ space group) used in this study. The coupling of magnetisation to an external magnetic field $\mathbf{H}$ is defined by the Zeeman energy density term $w_\text{z} = - \mu_{0}M_\text{s}\mathbf{H} \cdot \mathbf{m}$, with $\mu_{0}$ being the magnetic constant. The $w_\text{d}$ term is the demagnetisation (magnetostatic) energy density. Because $w_\text{d}$ is crucial for the stability of skyrmionic states in confined helimagnetic nanostructures,~\cite{Beg2015} we include its contribution in all subsequent simulations. The last term is the magnetocrystalline anisotropy energy density $w_\text{a}$, and because it does not play an important role in the stability of skyrmionic states in the studied system,~\cite{Beg2015} we assume the simulated material is isotropic and neglect the magnetocrystalline anisotropy energy contribution. The FeGe material parameters we use are the following:~\cite{Beg2015} saturation magnetisation $M_\text{s} = 384 \,\text{kA}\,\text{m}^{-1}$, exchange energy constant $A = 8.78 \,\text{pJ}\,\text{m}^{-1}$, and Dzyaloshinskii-Moriya energy constant $D = 1.58 \,\text{mJ}\,\text{m}^{-2}$. In our model, we do not assume any translational invariance of magnetisation in the out-of-film direction, which significantly changes the energy landscape both in infinitely large thin films~\cite{Rybakov2013} and in confined thin film nanostructures.~\cite{Beg2015} The relevant length scales in the simulated system are the exchange length $l_\text{ex} = \sqrt{2A/\mu_{0}M_\text{s}^{2}} = 9.73 \,\text{nm}$ and helical length $L_\text{D} = 4\pi A/D = 70 \,\text{nm}$. We choose the finite element mesh discretisation so that the maximum spacing between two neighbouring mesh nodes is below $l_\text{max} = 3 \,\text{nm}$, which is significantly smaller than the exchange length $l_\text{ex}$.

The magnetisation dynamics is governed by the Landau-Lifshitz-Gilbert (LLG) equation~\cite{Landau1935, Gilbert2004}
\begin{equation}
  \label{eq:llg_equation}
  \frac{\partial \mathbf{m}}{\partial t} = -\gamma_{0}^{*} \mathbf{m} \times \mathbf{H}_\text{eff} + \alpha\mathbf{m} \times \frac{\partial \mathbf{m}}{\partial t},
\end{equation}
where $\gamma_{0}^{*} = \gamma_{0} (1+\alpha^{2})$, with $\gamma_{0} = 2.21 \times 10^{5} \,\text{m}\,\text{A}^{-1}\text{s}^{-1}$ and $\alpha \ge 0$ is the Gilbert damping. We compute the effective magnetic field $\mathbf{H}_\text{eff}$ using
\begin{equation}
  \mathbf{H}_\text{eff} = -\frac{1}{\mu_{0}M_\text{s}} \frac{\delta E[\mathbf{m}]}{\delta \mathbf{m}},
\end{equation}
where $E[\mathbf{m}]$ is the total magnetic energy functional, given by Eq.~(\ref{eq:total_energy}). We validated the boundary conditions by running a series of simulations and reproducing the results reported by Rohart and Thiaville~\cite{Rohart2013}.

We implemented the presented model in the finite element method framework and developed a micromagnetic simulation tool Finmag (successor of Nmag~\cite{Fischbacher2007}). For the low-level finite element operations, we use FEniCS project~\cite{Logg2012} and for the adaptive step time integration we use Sundials/CVODE solver.~\cite{Hindmarsh2005, Cohen1996} For visualisation, we use Matplotlib~\cite{Hunter2007} and ParaView.~\cite{Ahrens2005}

We study the dynamic properties of skyrmionic states using two different methods: eigenvalue method~\cite{DAquino2009} and ringdown method.~\cite{McMichael2005} In both eigenvalue and ringdown methods, we firstly compute an equilibrium magnetisation configuration $\mathbf{m}_{0}$ by integrating a set of dissipative time-dependent equations, starting from a specific initial magnetisation configuration, until the condition of vanishing torque ($\mathbf{m} \times \mathbf{H}_\text{eff}$) is satisfied. The details on selecting the initial magnetisation configurations can be found in Ref.~\onlinecite{Beg2015}. We perform all relaxations in this work down to the maximum precision limited by the unavoidable numerical noise. Because the magnetisation dynamics is not of interest in the relaxation process, we set the Gilbert damping in this stage to $\alpha = 1$.

We perform the eigenvalue method computations in a finite element framework, motivated by the analytic procedure by d'Aquino~\textit{et al.}~\cite{DAquino2009} The perturbation of the system's magnetisation from its equilibrium state $\mathbf{m}_{0}$ can be written as $\mathbf{m}(t) = \mathbf{m}_{0} + \varepsilon \mathbf{v}(t)$, where $\varepsilon \in \mathbb{R}^{+}$ and $\mathbf{v}(t) \perp \mathbf{m}_{0}$ because of the imposed micromagnetic condition $|\mathbf{m}| = 1$. If this perturbation expression is inserted into the undamped LLG equation, we obtain
\begin{equation}
  \label{eq:eigenvalue_method1}
  \frac{\partial}{\partial t} (\mathbf{m}_{0} + \varepsilon \mathbf{v}(t)) = -\gamma_{0}^{*} (\mathbf{m}_{0} + \varepsilon \mathbf{v}(t)) \times \mathbf{H}_\text{eff}(\mathbf{m}_{0} + \varepsilon \mathbf{v}(t)).
\end{equation}
By using a Taylor expansion $\mathbf{H}_\text{eff}(\mathbf{m}_{0} + \varepsilon \mathbf{v}(t)) = \mathbf{H}_{0} + \varepsilon\mathbf{H}_\text{eff}'(\mathbf{m}_{0}) \cdot \mathbf{v}(t) + \mathcal{O}(\varepsilon^{2})$, where $\mathbf{H}_{0} = \mathbf{H}_\text{eff}(\mathbf{m}_{0})$, and knowing that $\partial \mathbf{m}_{0} / \partial t = 0$ and $\mathbf{m}_{0} \times \mathbf{H}_{0} = 0$, we get
\begin{equation}
  \label{eq:eigenvalue_method2}
  \frac{\partial}{\partial t} \mathbf{v}(t) = -\gamma_{0}^{*} \left[\mathbf{v}(t) \times \mathbf{H}_{0} +  \mathbf{m}_{0} \times (\mathbf{H}_\text{eff}'(\mathbf{m}_{0}) \cdot \mathbf{v}(t))\right],
\end{equation}
where all $\mathcal{O}(\varepsilon^{2})$ terms and higher are neglected. When the system is in its equilibrium, because $\mathbf{H}_\text{eff}(\mathbf{m}_{0}) \parallel \mathbf{m}_{0}$ and $|\mathbf{m}_{0}| = 1$, the equilibrium effective field can be written as $\mathbf{H}_{0} = h_{0}\mathbf{m}_{0}$, where $h_{0} = |\mathbf{H}_{0}|$. Now, if all vector fields are discretised on the finite elements mesh, Eq.~(\ref{eq:eigenvalue_method2}) becomes
\begin{equation}
  \label{eq:eigenvalue_method3}
  \frac{\partial}{\partial t} \mathbf{v}(t) = \gamma_{0}^{*} \mathbf{m}_{0} \times \left[(h_{0} \mathbb{1} - \mathbf{H}_\text{eff}'(\mathbf{m}_{0})) \cdot \mathbf{v}(t)\right].
\end{equation}
Using the matrix $\Lambda(\mathbf{m}_{0})$ with property $\mathbf{m}_{0} \times \mathbf{x} = \Lambda(\mathbf{m}_{0}) \cdot \mathbf{x}$, Eq.~(\ref{eq:eigenvalue_method3}) can be written as
\begin{equation}
  \label{eq:eigenvalue_method4}
  \frac{\partial}{\partial t} \mathbf{v}(t) = A \cdot \mathbf{v}(t),
\end{equation}
where $A = \gamma_{0}^{*} \Lambda(\mathbf{m}_{0}) \left[h_{0} \mathbb{1} - \mathbf{H}_\text{eff}'(\mathbf{m}_{0})\right]$. This linear differential equation has a full set of solutions that can be expressed as $\mathbf{v}(t) = \tilde{\mathbf{v}} \text{e}^{i2\pi ft}$, where $\tilde{\mathbf{v}}$ is a constant vector field. Using this ansatz, Eq.~(\ref{eq:eigenvalue_method4}) becomes the eigenvalue problem
\begin{equation}
  \label{eq:eigenvalue_method5}
  i2\pi f \tilde{\mathbf{v}} = A \tilde{\mathbf{v}}.
\end{equation}
We solve this eigenvalue problem using Python wrappers for the ARPACK~\cite{Maschho1996} solvers which are implemented in the SciPy~\cite{Oliphant2007, Walt2011} package, which results in a set of resonant frequencies $f$ and eigenvectors $\tilde{\mathbf{v}}$ from which we express the magnetisation dynamics as $\mathbf{m}(t) = \mathbf{m}_{0} + \tilde{\mathbf{v}}\text{e}^{i2\pi ft}$.

In the ringdown method, similar to the eigenvalue method, we firstly relax the system to its equilibrium magnetisation state $\mathbf{m}_{0}$. After that, we perturb the system from its equilibrium by applying a time-dependent $\mathbf{h}(t) = h_\text{max} \operatorname{sinc}(2\pi f_\text{c}t) \hat{\mathbf{e}}$ external magnetic field excitation~\cite{Kim2014, Venkat2013} over $t_\text{exc}=0.5 \,\text{ns}$, where $h_\text{max}$ is the maximum excitation field value, $f_\text{c} = 100 \,\text{GHz}$ is the cut-off frequency, $\hat{\mathbf{e}}$ is the direction in which the excitation is applied, and $\operatorname{sinc}(2\pi f_\text{c}t)$ is the unnormalised cardinal sine function
\begin{equation}
  \label{eq:sinc_pulse}
  \operatorname{sinc}(2\pi f_\text{c}t) =
  \begin{cases}
    \frac{\sin (2\pi f_\text{c}t)}{2\pi f_\text{c}t}, & \text{for }t \neq 0 \\
    1, & \text{for } t = 0. \\
  \end{cases}
\end{equation}
The time-dependence of the used excitation $h(t)$ is shown in Fig.~\ref{fig:methods}(b). Computing the Fourier transform of $h(t)$ shows that using this excitation enables us to excite all eigenmodes (which are allowed by the direction of excitation $\hat{\mathbf{e}}$) in the $[0, f_\text{c}]$ range approximately equally, as demonstrated in Fig.~\ref{fig:methods}(c). We compute the $h_\text{max}$ value so that $H^{f}=0.5 \,\text{mT}$ is the excitation amplitude at any frequency.~\cite{Kim2014} More precisely, the maximum value of the cardinal sine wave excitation is $h_\text{max} = 2f_\text{c}t_\text{exc}H^{f} = 50 \,\text{mT}$. We apply the excitation in two experimentally feasible directions: (i) in-plane $\hat{\mathbf{e}} = \hat{\mathbf{x}}$ and (ii) out-of-plane $\hat{\mathbf{e}} = \hat{\mathbf{z}}$. After the system is perturbed from its equilibrium state, we simulate the magnetisation dynamics for $t_\text{sim} = 20 \,\text{ns}$ and sample the magnetisation field $\mathbf{m}(\mathbf{r}_{i}, t_{j})$ at all mesh nodes $\mathbf{r}_{i}$ at uniform time steps $t_{j} = j\Delta t$ ($\Delta t = 5 \,\text{ps}$). Although the excitation is sufficiently small so that the perturbation from the equilibrium state can be approximated linearly, in order to make sure we do not introduce any nonlinearities to the system's dynamics with the excitation, we delay sampling by $2 \,\text{ns}$ after the excitation field is removed.

Finally, we analyse the recorded magnetisation dynamics $\mathbf{m}(\mathbf{r}_{i}, t_{j})$ using: (i) spatially averaged and (ii) spatially resolved methods.~\cite{Baker2017} We subtract the time-independent equilibrium magnetisation configuration $\mathbf{m}_{0}(\mathbf{r}_{i})$ from the recorded magnetisation dynamics and perform the Fourier analysis only on the time-dependent part $\Delta \mathbf{m}(\mathbf{r}_{i}, t_{j}) = \mathbf{m}(\mathbf{r}_{i}, t_{j}) - \mathbf{m}_{0}(\mathbf{r}_{i})$. In the spatially averaged analysis, we compute all three spatially averaged magnetisation components $\left< \Delta {m}_{k}(t_{j})\right>$, $k = x, y, z$, at all time steps $t_{j}$. After that, we apply a discrete Fourier transform and sum the squared Fourier coefficient moduli (which are proportional to the power) to obtain the power spectral density (PSD):
\begin{equation}
  P_\text{sa}(f) = \sum_{k=x,y,z} \bigg| \sum_{j=1}^{n} \left< \Delta {m}_{k}(t_{j})\right> \text{e}^{-i2\pi ft_{j}} \bigg|^{2},
\end{equation}
where $n$ is the number of time steps at which the magnetisation dynamics was sampled. On the other hand, in the spatially resolved analysis, we firstly compute the discrete Fourier transform at all mesh nodes (separately for all three magnetisation components) and then compute the PSD as the spatial average of the squared Fourier coefficient moduli:~\cite{McMichael2005}
\begin{equation}
  \label{eq:spatially_resolved}
  P_\text{sr}(f) = \sum_{k=x,y,z} \frac{1}{N}\sum_{i=1}^{N}\bigg| \sum_{j=1}^{n} m_{k}(\mathbf{r}_{i}, t_{j})\text{e}^{-i2\pi ft_{j}} \bigg|^{2},
\end{equation}
where $N$ is the number of finite element mesh nodes. Because the power values in PSD are in arbitrary units (a.u.), we normalise all PSDs in this work so that $\int_{0}^{f_\text{max}} P(f) \,\text{d}f = 1$, where $f_\text{max} = 50 \,\text{GHz}$, and show them in the logarithmic scale. We choose to analyse dynamics of skyrmionic states in the $[0, 50 \,\text{GHz}]$ range in order to avoid the presence of artefact peaks in PSDs due to aliasing~\cite{Antoniou2005} as a consequence of discrete time sampling limitations. Although, the frequency resolution in the eigenvalue method is determined by the machine precision, the frequency resolution for the ringdown method is $\Delta f = (n\Delta t)^{-1} \approx {t_\text{sim}}^{-1} = 0.05 \,\text{GHz}$, where $n = t_\text{sim}/\Delta t + 1$ is the number of sampling points during the sampling simulation stage.

\section{Results}
We study the dynamics of all three equilibrium skyrmionic states that can be observed at zero external magnetic field in confined thin film helimagnetic disk samples with diameters $d \le 180 \,\text{nm}$. More precisely, we explore the resonance frequencies and corresponding eigenmode magnetisation dynamics of the ground state incomplete skyrmion (iSk) and isolated skyrmion (Sk) states, as well as the metastable target (T) configuration. The difference between these states is in how many times the magnetisation configuration covers the sphere. A quantity that is usually used to determine whether a magnetisation configuration covers the sphere is the skyrmion number~\cite{Heinze2011} $S$. However, in confined helimagnetic nanostructures, an additional tilting of magnetisation at the sample edges~\cite{Rohart2013} in the winding direction opposite to the skyrmion configuration reduces the absolute skyrmion number value~\cite{Sampaio2013, Beg2015}. This does not allow us to determine what skyrmionic state is present is the sample because $|S| < 1$ for all of them. To address this, a scalar value~\cite{Beg2015} $S_\text{a}$ in a three-dimensional sample is defined as
\begin{equation}
  \label{eq:skyrmion_number_3d}
  S_\text{a} = \frac{1}{8\pi} \int \left|\mathbf{m} \cdot \left( \frac{\partial \mathbf{m}}{\partial x} \times \frac{\partial \mathbf{m}}{\partial y}\right) \right| \,\text{d}^{3}r.
\end{equation}
This scalar value has no mathematical or physical interpretation, and is defined merely to support the classification of skyrmionic states in confined nanostructures.

Using the eigenvalue method, we find all existing eigenmodes by computing their resonance frequencies and magnetisation dynamics. However, this method does not allow us to determine what eigenmodes can be excited using a particular excitation. Therefore, we employ the ringdown method for an in-plane and an out-of-plane excitation and overlay the resulting spatially averaged and spatially resolved power spectral densities (PSDs) with the resonance frequencies obtained from the eigenvalue method. If the eigenvalue method resonance frequency coincides with a PSD peak, this implies that the corresponding eigenmode can be ``activated'' using a specific excitation and we mark it using a triangle ($\triangle$) symbol. All other eigenmodes, that cannot be activated using a particular excitation, we mark with a circle ($\circ$) symbol. Throughout this work, we study the magnetisation dynamics below $50 \,\text{GHz}$. We analyse the target state dynamics in Supplementary Section S1.

\begin{figure*}
  \includegraphics{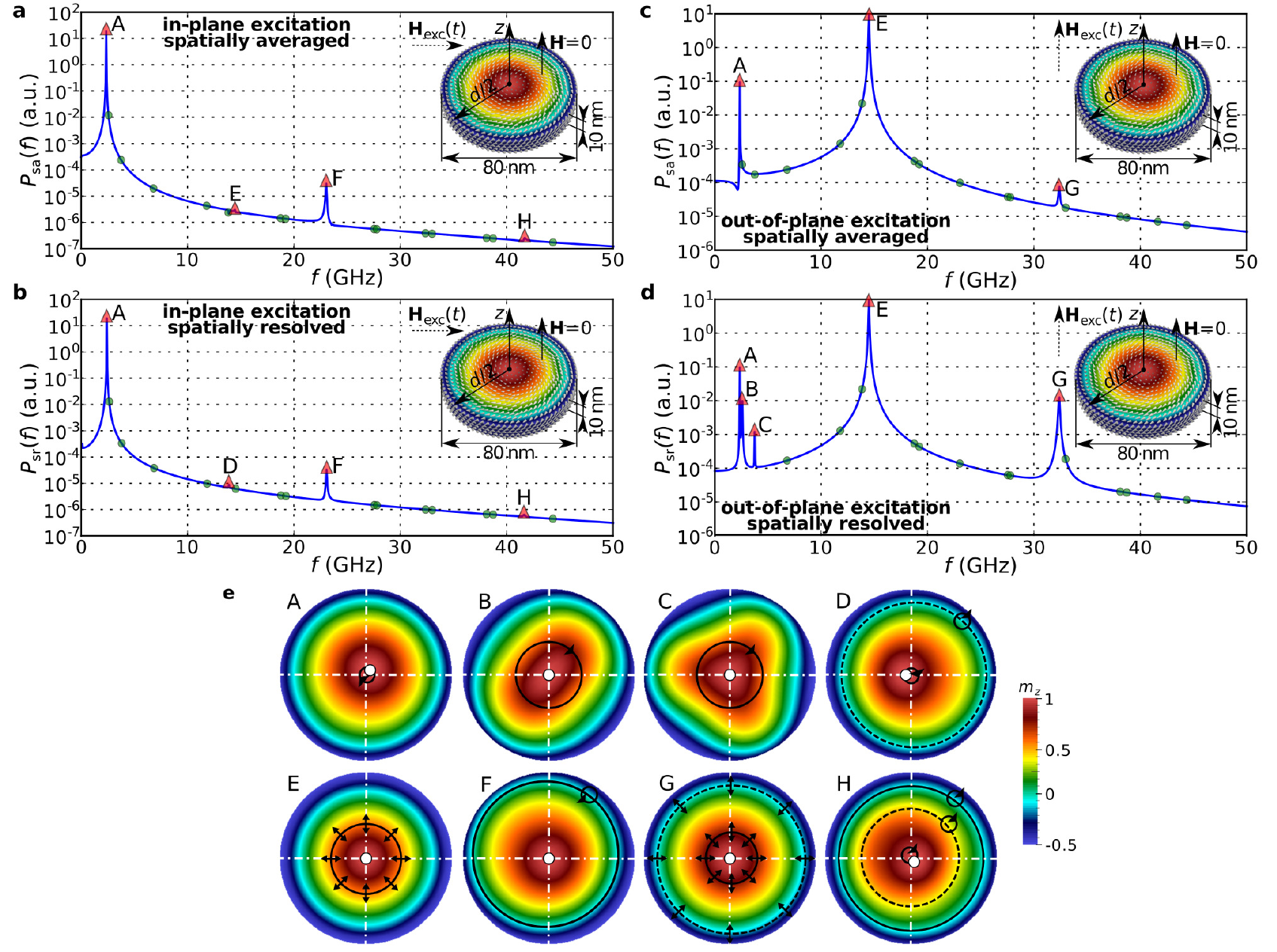}
  \caption{\label{fig:isk_psd} The power spectral densities (PSDs) of an incomplete skyrmion (iSk) ground state at zero external magnetic field in a $80 \,\text{nm}$ diameter FeGe disk sample with $10 \,\text{nm}$ thickness. (a)~Spatially averaged and (b)~spatially resolved PSDs for an in-plane excitation, together with overlaid resonance frequencies computed using the eigenvalue method. The resonant frequencies obtained using the eigenvalue method are marked with a triangle symbol ($\triangle$) if they can be activated using a particular excitation and with a circle symbol ($\circ$) otherwise. (c)~Spatially averaged and (d)~spatially resolved PSDs for an out-of-plane excitation. (e)~Schematic representations of magnetisation dynamics associated with the identified eigenmodes. Schematically, we represent the skyrmionic state core with a circle symbol, together with a directed loop if it gyrates around its equilibrium position. Contour rings represented using dashed lines revolve/breathe out-of-phase with respect to the those marked with solid lines. The magnetisation dynamics animations of all identified eigenmodes are provided in Supplementary Video~1.}
\end{figure*}

\begin{figure*}
  \includegraphics{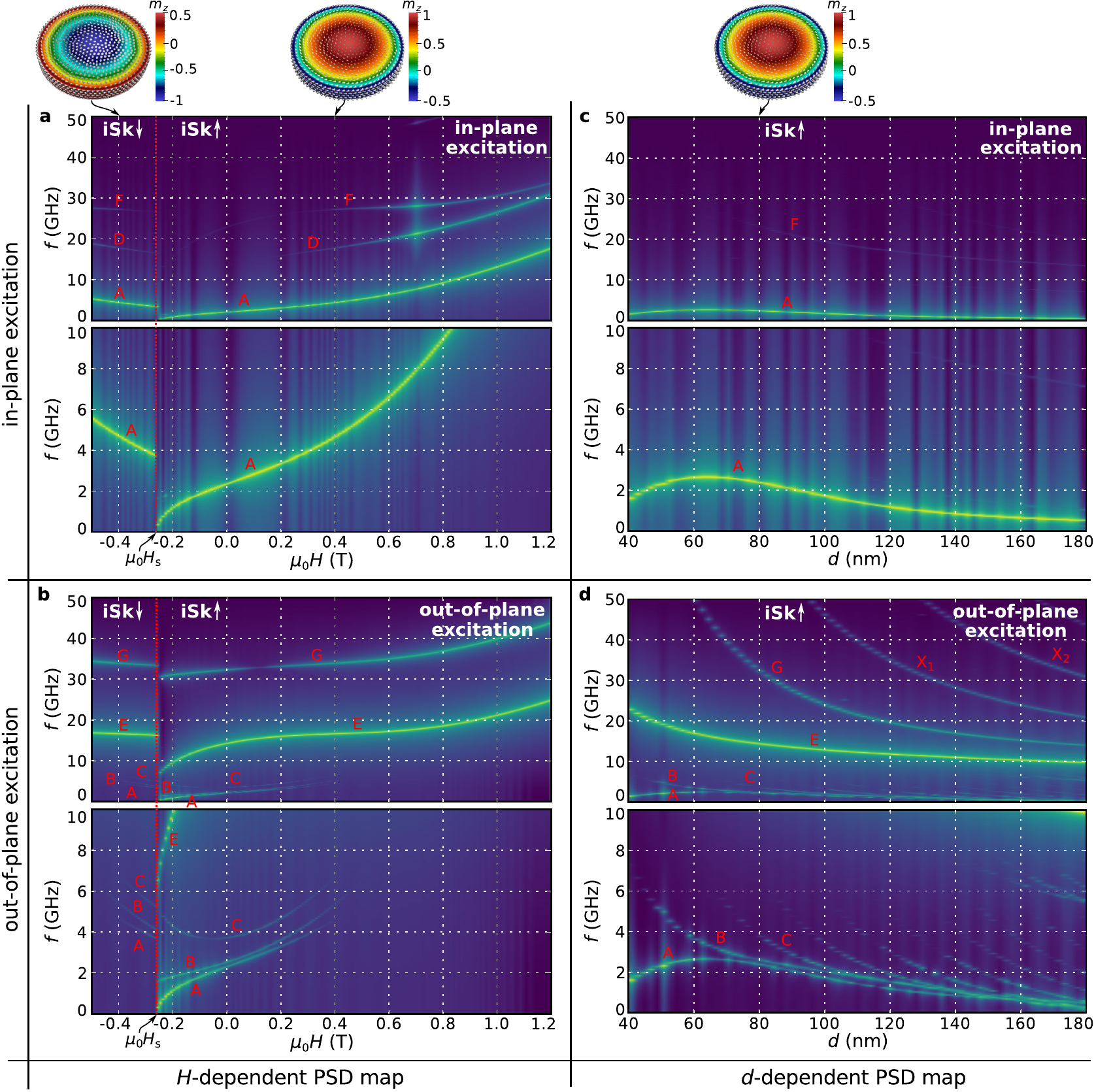}
  \caption{\label{fig:isk_sweep} Power spectral density (PSD) maps showing the dependence of incomplete skyrmion (iSk) state resonant frequencies on the external magnetic bias field changed between $-0.5 \,\text{T}$ and $1.2 \,\text{T}$ in steps of $10 \,\text{mT}$ for $d = 80 \,\text{nm}$ when the system is excited using (a)~in-plane and (b)~out-of-plane excitation. The dependence of resonance frequencies on the disk sample diameter varied between $40 \,\text{nm}$ and $180 \,\text{nm}$ in steps of $2 \,\text{nm}$ at zero external magnetic field for (c)~in-plane and (d)~out-of-plane excitation. We show two plots for every PSD map: one for the complete studied frequency range ($0-50 \,\text{GHz}$) and another plot in order to better resolve the low-frequency ($0-10 \,\text{GHz}$) part of the PSD map.}
\end{figure*}

\subsection{Incomplete skyrmion (iSk) state}\label{sec:isk}
The first magnetisation configuration that we study is the incomplete skyrmion (iSk) state. The magnetisation component $m_{z}$ of the iSk state, along the sample diameter, does not cover the whole $[-1, 1]$ range, which is required for the skyrmion configuration to be present in the sample, and because of that, the scalar value $S_{a}$ is in the $[0, 1]$ range.~\cite{Beg2015} In other works, this state is called either the quasi-ferromagnetic~\cite{Sampaio2013, Rohart2013} or the edged vortex~\cite{Du2013, Du2013a} state. The incomplete skyrmion state is in an equilibrium for all studied disk sample diameters $40 \,\text{nm} \le d \le 180 \,\text{nm}$ and at all external magnetic bias field values.~\cite{Beg2015} We explore the resonance frequencies and corresponding eigenmode magnetisation dynamics in a $80 \,\text{nm}$ diameter disk sample at zero external magnetic field, where the iSk state is not only in an equilibrium, but is also the ground state~\cite{Beg2015} (global energy minimum).

Firstly, we compute all existing eigenmodes using the eigenvalue method and show the magnetisation dynamics of all identified eigenmodes in Supplementary Video~1, and their schematic representations in Supplementary Section S3. Then, we excite the system using an in-plane excitation and show the spatially averaged (SA) and spatially resolved (SR) power spectral densities (PSDs) overlaid with the eigenvalue method resonant frequencies in Figs.~\ref{fig:isk_psd}(a) and~\ref{fig:isk_psd}(b), respectively. In these two PSDs, we identify five peaks (A, D, E, F, and H) and schematically represent their corresponding eigenmode magnetisation dynamics in Fig.~\ref{fig:isk_psd}(e). The lowest frequency and the most dominant eigenmode A at $2.35\,\text{GHz}$ consists of a dislocated incomplete skyrmion state core (where $m_{z} = 1$) revolving (gyrating) around its equilibrium position in the clockwise (CW) direction. Schematically, we represent the skyrmionic state core with a circle symbol, together with a directed loop if it gyrates around its equilibrium position. Consequently, we classify the eigenmode A as the gyrotropic (translational) mode. The eigenmode F at $23.04 \,\text{GHz}$ is the second most dominant eigenmode. Its magnetisation dynamics consists of a ring contour, defined by the constant magnetisation $z$~component distribution, revolving around the sample centre in the counterclockwise (CCW) direction. This eigenmode is not gyrotropic because the iSk state core remains at its equilibrium position. The eigenmode H at $41.65 \,\text{GHz}$, present in both SA and SR PSDs, is composed of the iSk state core together with two $m_{z}$ contour rings revolving in the CW direction. However, the inner contour revolves out-of-phase with respect to both the outer contour and the iSk state core. Because of that, we depict the inner contour ring using a dashed line and both the iSk state core loop and the outer contour ring using a solid line as a way of visualising the mutually out-of-phase dynamics. The eigenmode D is present only in the SR PSD at $13.83 \,\text{GHz}$ and consists of the iSk state core and a contour ring revolving in the CW direction, but mutually out-of-phase. So far, all identified eigenmodes are lateral, but in the SA PSD at $14.49 \,\text{GHz}$, we also identify a very weak eigenmode E with radially symmetric magnetisation dynamics. Although we expect that all eigenmodes present in the SA PSD are also present in the SR PSD, this is not the case for eigenmode E. We believe this is the case because this breathing eigenmode can be excited with an out-of-plane excitation, but emerges in simulations with an in-plane excitation due to the numerical noise, which is consistent with its small amplitude. This eigenmode, together with other breathing eigenmodes, will be discussed subsequently when we excite the iSk state using an out-of-plane excitation.

Now, we perturb the incomplete skyrmion state from its equilibrium using an out-of-plane excitation and show the spatially averaged (SA) and spatially resolved (SR) power spectral densities (PSDs) in Figs.~\ref{fig:isk_psd}(c) and~\ref{fig:isk_psd}(d), respectively. Using this excitation, we identify five eigenmodes (A, B, C, E, and G) and schematically represent their magnetisation dynamics in Fig.~\ref{fig:isk_psd}(e). The most dominant eigenmode E is present in both SA and SR PSDs at $13.83 \,\text{GHz}$. Its magnetisation dynamics consists of a $m_{z}$ contour ring that shrinks and expands periodically, while the overall magnetisation configuration remains radially symmetric. Because of that, we classify this eigenmode as a breathing mode. The second most dominant eigenmode is the gyrotropic mode A, which was also observed when the system was excited using an in-plane excitation, suggesting that it can be experimentally detected independent of the used excitation direction. The last eigenmode G present in both SA and SR PSDs at $32.37 \,\text{GHz}$ consists of two contour rings breathing mutually out-of-phase. More precisely, when one contour shrinks, another one expands, and vice versa. We schematically illustrate this out-of-phase breathing using dashed and solid lines depicting the contours. Finally, the eigenmodes B and C, visible only in the SR PSD at $2.57 \,\text{GHz}$ and $3.76 \,\text{GHz}$, respectively, can be understood as a particular magnetisation configuration rotating in the sample in the CW direction without dislocating their core, as shown in Fig.~\ref{fig:isk_psd}(e).

After analysing the incomplete skyrmion power spectral densities for $d=80 \,\text{nm}$ and $H=0$, we now explore how the resonance frequencies depend on the external magnetic bias field $H$ and the disk sample diameter $d$ for both an in-plane and an out-of-plane excitation. Firstly, we fix the disk sample diameter at $80 \,\text{nm}$ and reduce the external magnetic field from $1.2 \,\text{T}$ to $-0.5 \,\text{T}$ in steps of $10 \,\text{mT}$. More precisely, we start at $\mu_{0}H=1.2 \,\text{T}$ initialising the system using incomplete skyrmion configuration with positive core orientation iSk$\uparrow$, relax the system to its equilibrium, and then run both eigenvalue and ringdown simulations. After that, we reduce the external magnetic field by $10 \,\text{mT}$, relax the system to its equilibrium using the relaxed (equilibrium) state from the previous simulation as initial configuration, and run dynamics simulations. We iterate until we reach $-0.5 \,\text{T}$. We show the resulting $H$-dependent power spectral density (PSD) maps for an in-plane and an out-of-plane excitation in Figs.~\ref{fig:isk_sweep}(a) and \ref{fig:isk_sweep}(b), respectively. In these $H$-dependent PSD maps, a discontinuity in resonance frequencies at $-0.26 \,\text{T}$ is evident. This is the case because for $d=80 \,\text{nm}$ and $-0.26 \,\text{T} \le \mu_{0}H \le 1.2 \,\text{T}$, the iSk state with positive ($m_{z} = 1$) core orientation (iSk$\uparrow$) is in an equilibrium. However, for $\mu_{0}H < -0.26 \,\text{T}$, the iSk$\uparrow$ is not in an equilibrium anymore and the iSk state reverses its orientation to the negative ($m_{z}=-1$) direction (iSk$\downarrow$) in order to reduce its Zeeman energy. This is consistent with the incomplete skyrmion hysteretic behaviour studies.~\cite{Beg2015} Secondly, we change the disk sample diameter $d$ between $40 \,\text{nm}$ and $180 \,\text{nm}$ in steps of $2 \,\text{nm}$ at zero external magnetic bias field and show in Figs.~\ref{fig:isk_sweep}(c) and \ref{fig:isk_sweep}(d) the $d$-dependent PSD maps for an in-plane and an out-of-plane excitation, respectively. In PSD maps, we show the spatially resolved PSDs, computed using Eq.~(\ref{eq:spatially_resolved}), because in comparison to the spatially averaged PSDs, they exhibit more resonance peaks.~\cite{McMichael2005} We show two plots for every PSD map: one for the complete studied frequency range ($0-50 \,\text{GHz}$) and another plot in order to better resolve the low-frequency ($0-10 \,\text{GHz}$) part of the PSD map.

In the case of an in-plane excitation, three lateral eigenmodes (A, D, and F) are visible in the $H$-dependent PSD map, shown in Fig.~\ref{fig:isk_sweep}(a), and in the iSk$\uparrow$ range their frequencies nonlinearly and monotonically increase with $H$. Eigenmodes D and F are not as dominant as eigenmode A in the PSD map below approximately $0.3 \,\text{T}$, which results in the lack of sufficient contrast for them to be visible. Now, if we change the direction of excitation, five eigenmodes (A, B, C, E, and G) are visible in the $H$-dependent PSD map, as shown in Fig.~\ref{fig:isk_sweep}(b). In the iSk$\uparrow$ range, eigenmodes A, B, and C are visible only between $H_\text{s}$ and approximately $0.4 \,\text{T}$. In Fig.~\ref{fig:isk_psd}(d) at zero external magnetic field, eigenmodes A, B, and C have very similar frequencies which makes it difficult for experimentalists to determine which eigenmode the resonance frequency they measure belongs to. From the out-of-plane $H$-dependent PSD map in Fig.~\ref{fig:isk_sweep}(b), we observe that this difficulty can be resolved by reducing the external magnetic field towards the switching field. More precisely, the frequencies of eigenmodes A and B both decrease, but only the frequency of eigenmode A approaches zero. On the contrary, the frequency of eigenmode C increases by reducing $H$. Also, the separation between A, B, and C peaks in the PSD increases, by either increasing or decreasing external magnetic field value. We show the dependence of their frequencies at high external magnetic fields in Supplementary Section~S2 using the eigenvalue method and demonstrate that no eigenmode crossing occurs. In addition, as we previously discussed, by changing the excitation to an in-plane direction, only eigenmode A would be present. The eigenmodes E and G are visible in the whole examined range of $H$, and their frequencies increase nonlinearly and monotonically with external magnetic bias field. Interestingly, the frequency of eigenmode A approaches zero near the switching field $\mu_{0}H_\text{s} = -0.26 \,\text{T}$, suggesting that this gyrotropic eigenmode might be the reversal (zero) mode of the incomplete skyrmion state in the studied sample.

By varying the disk sample diameter $d$, for an in-plane excitation, we observe the gyrotropic eigenmode A frequency increasing between $40 \,\text{nm}$ and $64 \,\text{nm}$ (where it reaches its maximum), and then decreasing with $d$, as shown in Fig.~\ref{fig:isk_sweep}(c). Another visible eigenmode in the PSD map above approximately $74 \,\text{nm}$, for an in-plane excitation, is the eigenmode F whose frequency monotonically decreases with $d$. In the case of an out-of-plane excitation, we identify seven (A, B, C, E, G, X$_{1}$, and X$_{2}$) eigenmodes in the PSD map shown in Fig.~\ref{fig:isk_sweep}(d). The magnetisation dynamics of all these eigenmodes was discussed before, except X$_{1}$ and X$_{2}$, because they were not present in the PSDs below the maximum studied frequency $50 \,\text{GHz}$ for $d=80 \,\text{nm}$. The eigenmode A frequency dependence is the same as for an in-plane excitation and another six eigenmodes (B, C, E, G, X$_{1}$, and X$_{2}$) frequencies monotonically decrease with the disk sample diameter.

\subsection{Isolated skyrmion (Sk) state}\label{sec:sk}
In this section, we explore the dynamics of an isolated skyrmion (Sk) state, for which the magnetisation $z$ component covers the $[-1, 1]$ range once (plus the additional magnetisation tilting at the boundaries due to the specific boundary conditions~\cite{Rohart2013}) along the disk sample diameter, and consequently, the scalar value $S_\text{a}$ is in the $[1, 2]$ range.~\cite{Beg2015} The Sk state is in an equilibrium~\cite{Beg2015} for $d \ge 70 \,\text{nm}$ and $\mu_{0}H \le 1.1 \,\text{T}$. We study the Sk state dynamics for a $150 \,\text{nm}$ diameter disk sample at zero external magnetic bias field, where the Sk state is not only in an equilibrium, but is also the ground state.~\cite{Beg2015}

\begin{figure*}
  \includegraphics{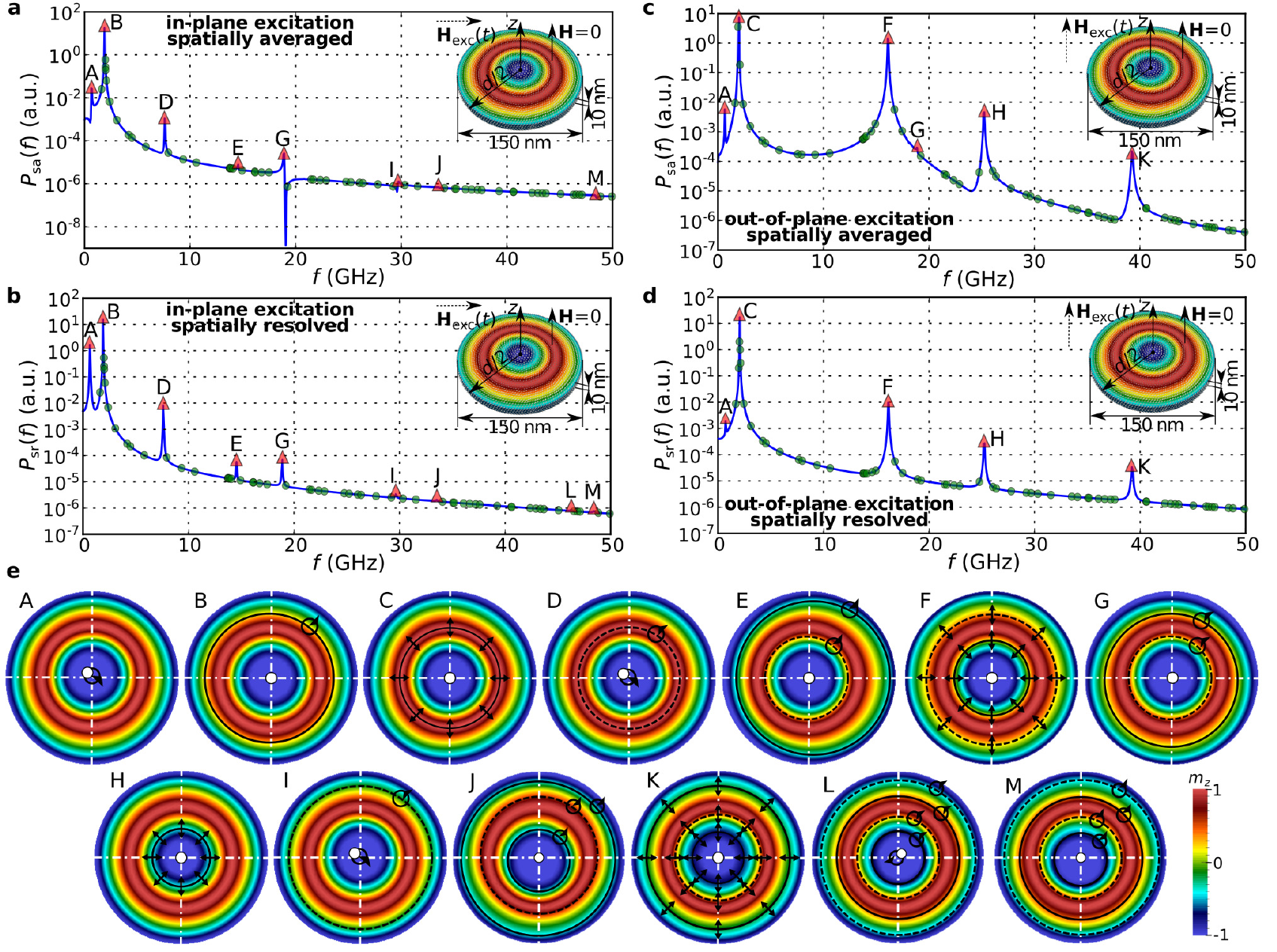}
  \caption{\label{fig:sk_psd} The power spectral densities (PSDs) of an isolated skyrmion (Sk) ground state in a $150 \,\text{nm}$ diameter FeGe disk sample with $10 \,\text{nm}$ thickness at zero external magnetic bias field. (a)~Spatially averaged and (b)~spatially resolved PSDs for an in-plane excitation, together with overlaid resonance frequencies computed using the eigenvalue method. The resonant frequencies obtained using the eigenvalue method are marked with a triangle symbol ($\triangle$) if they can be activated using a particular excitation and with a circle symbol ($\circ$) otherwise. (c)~Spatially averaged and (d)~spatially resolved PSDs computed when the Sk state is perturbed from its equilibrium with an out-of-plane excitation. (e)~Schematic representations of magnetisation dynamics associated with the identified eigenmodes. Schematically, we represent the skyrmionic state core with a circle symbol, together with a directed loop if it gyrates around its equilibrium position. Contour rings represented using dashed lines revolve/breathe out-of-phase with respect to the those marked with solid lines. The magnetisation dynamics animations of all identified eigenmodes are provided in Supplementary Videos~2~and~3.}
\end{figure*}

\begin{figure*}
  \includegraphics{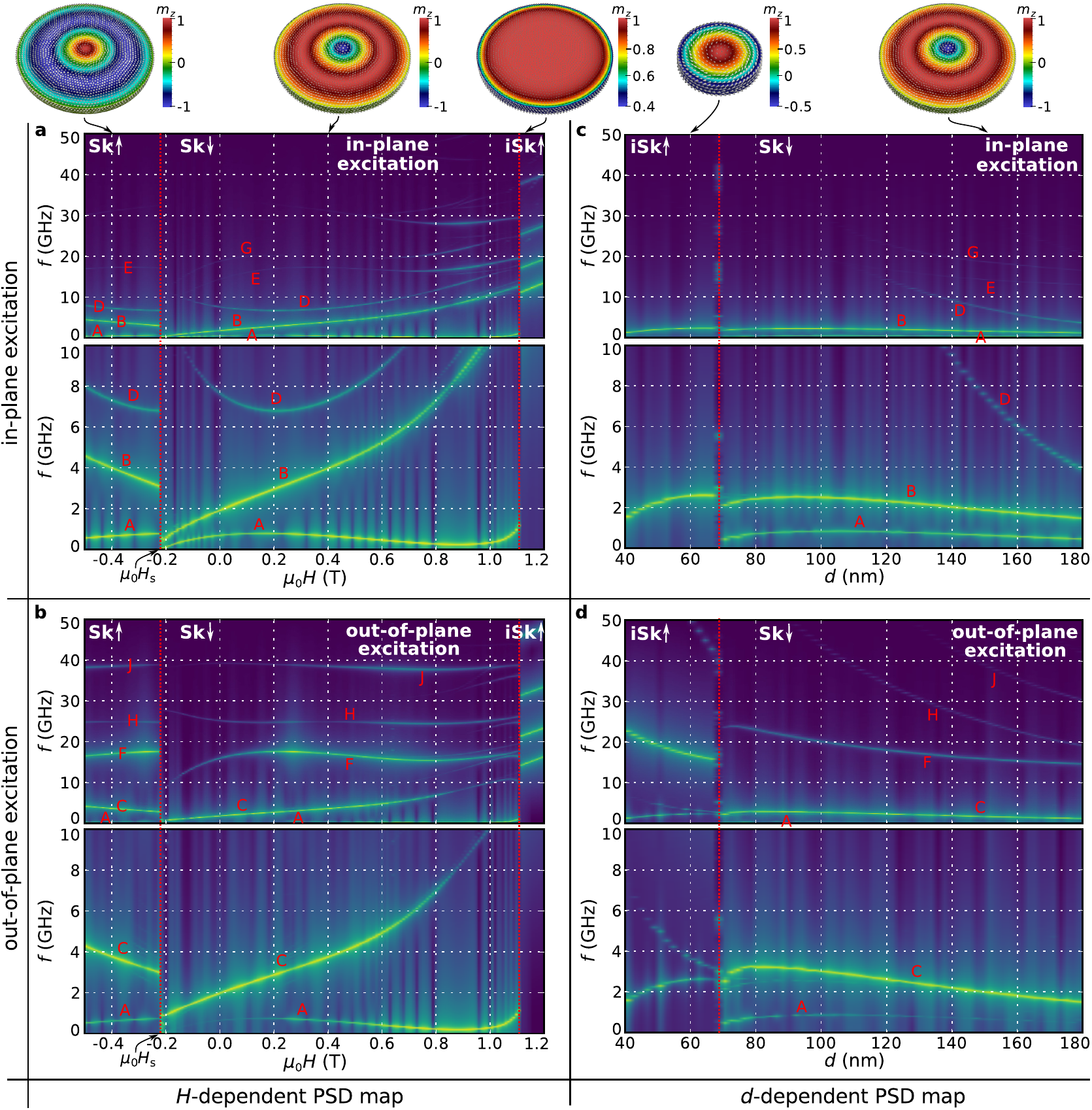}
  \caption{\label{fig:sk_sweep} Power spectral density (PSD) maps showing the dependence of isolated skyrmion (Sk) state resonant frequencies on the external magnetic bias field changed between $-0.5 \,\text{T}$ and $1.2 \,\text{T}$ in steps of $10 \,\text{mT}$ for $d = 150 \,\text{nm}$ when the system is excited using (a)~in-plane and (b)~out-of-plane excitation. The dependence of resonance frequencies on the disk sample diameter varied between $40 \,\text{nm}$ and $180 \,\text{nm}$ in steps of $2 \,\text{nm}$ at zero external magnetic field for (c)~in-plane and (d)~out-of-plane excitation. We show two plots for every PSD map: one for the complete studied frequency range ($0-50 \,\text{GHz}$) and another plot in order to better resolve the low-frequency ($0-10 \,\text{GHz}$) part of the PSD map.}
\end{figure*}

After we perform the eigenvalue method computations, we excite the system using an in-plane excitation and show the spatially averaged (SA) and spatially resolved (SR) power spectral densities (PSDs) in Figs.~\ref{fig:sk_psd}(a) and \ref{fig:sk_psd}(b), respectively. The magnetisation dynamics of all existing eigenmodes computed using the eigenvalue method are shown in Supplementary Videos~2~and~3, and their schematic representations in Supplementary Section~S3. In both SA and SR PSDs, we identify nine peaks (eigenmodes A, B, D, E, G, I, J, L, and M), and show their schematic representations in Fig.~\ref{fig:sk_psd}(e). The lowest frequency eigenmode at $0.67 \,\text{GHz}$ is the gyrotropic eigenmode A. Its magnetisation dynamics consists of a dislocated Sk state core (where $m_{z} = -1$) gyrating around its equilibrium position in the CCW direction. In both PSDs, the eigenmode B at $1.91 \,\text{GHz}$ is the most dominant one, and consists of a contour ring (defined as a constant magnetisation $z$ component distribution) revolving in the CW direction. The eigenmode D at $7.61 \,\text{GHz}$ is composed of both the Sk state core and a magnetisation contour ring revolving in the CCW direction, but mutually out-of-phase. At $14.54 \,\text{GHz}$, we identify an eigenmode E with two magnetisation contour rings revolving mutually out-of-phase in the CW direction. Similarly, the eigenmode G at $18.89 \,\text{GHz}$ also consists of two contour rings revolving mutually out-of-phase, but now in the CCW direction. The four remaining eigenmodes (I, J, L, and M) are significantly weaker in both PSDs when compared to the power of previously discussed eigenmodes. Their magnetisation dynamics, shown in Fig.~\ref{fig:sk_psd}(e), are all lateral and contain different combinations of revolving contour rings and revolving Sk state core.

Now, we change the excitation to be in the out-of-plane direction. The computed spatially averaged and spatially resolved power spectral densities, overlaid with the resonance frequencies obtained from the eigenvalue method, are shown in Figs.~\ref{fig:sk_psd}(c) and \ref{fig:sk_psd}(d), respectively. In this case, we observe five peaks (eigenmodes A, C, F, H, and K) in both PSDs, and a significantly weaker lateral eigenmode G (previously discussed) in SA PSD. We show the schematic representation of their magnetisation dynamics in Fig.~\ref{fig:sk_psd}(e). Similar to the incomplete skyrmion state, the gyrotropic eigenmode A can also be activated with an out-of-plane excitation. The lowest frequency breathing eigenmode C at $2.00 \,\text{GHz}$ consists of a single contour ring that shrinks and expands periodically. An eigenmode F at $16.12 \,\text{GHz}$ is composed of two contour rings breathing mutually out-of-phase. Similar to the eigenmode C, the eigenmode H at $25.22 \,\text{GHz}$ consists of a single breathing contour, but now with a smaller contour diameter (smaller $m_{z}$). At $39.25 \,\text{GHz}$, we identify the highest frequency breathing eigenmode K in the studied frequency range, which contains three breathing contours, where the inner and the outer contours breathe out-of-phase with respect to the middle one.

So far, we analysed the isolated skyrmion state dynamics for $d = 150 \,\text{nm}$ and $H=0$. Now, our objective is to determine how the resonance frequencies depend on disk sample diameter and external magnetic field. We compute the $H$-dependent PSD map in two parts in order to obtain how the resonance frequencies change in the entire range of external magnetic field values where the isolated skyrmion state with negative core orientation (Sk$\downarrow$) is in equilibrium. More precisely, the system initialised with Sk$\downarrow$ at high external magnetic fields relaxes to the incomplete skyrmion state with positive core orientation~\cite{Beg2015} iSk$\uparrow$. Consequently, if we keep reducing $H$ and use the equilibrium state from previous simulation iteration as initial state, we could not reach the Sk$\downarrow$ state. Therefore, we firstly fix the disk sample diameter to $150 \,\text{nm}$, set $H=0$, initialise the system using Sk$\downarrow$ configuration, relax the system, and run dynamics simulations. Then we increase the external magnetic field by $10 \,\text{mT}$ using the equilibrium state from previous simulation as the initial state at new value of external magnetic field. We iterate this until we reach $1.2 \,\text{T}$. Similarly, starting from zero external magnetic field, we reduce $\mu_{0}H$ in steps of $10 \,\text{mT}$, until $\mu_{0}H=-0.5 \,\text{T}$ is reached. We show the $H$-dependent PSD maps for an in-plane and an out-of-plane excitation in Fig.~\ref{fig:sk_sweep}(a) and \ref{fig:sk_sweep}(b), respectively. In these PSD maps, two discontinuities in resonant frequencies at $-0.24 \,\text{T}$ and $1.12 \,\text{T}$ are present. The first discontinuity occurs because decreasing $H$ causes the Sk state core with negative ($m_{z} = -1$) orientation (Sk$\downarrow$) to switch to the positive ($m_{z} = 1$) direction (Sk$\uparrow$) at the switching field $\mu_{0}H_\text{s} = -0.24 \,\text{T}$. On the other hand, the discontinuity at $1.12 \,\text{T}$ occurs because, above this value, the Sk$\downarrow$ is not in an equilibrium anymore and the system relaxes to the incomplete skyrmion state with positive core orientation (iSk$\uparrow$). Secondly, at $H=0$, we vary $d$ between $40 \,\text{nm}$ and $180 \,\text{nm}$ in steps of $2 \,\text{nm}$ and show the $d$-dependent PSD maps in Figs.~\ref{fig:sk_sweep}(c) and \ref{fig:sk_sweep}(d) for an in-plane and an out-of-plane excitation, respectively. Now, a single discontinuity in resonance frequencies is present at $70 \,\text{nm}$, below which the disk sample diameter becomes too small to accommodate the full magnetisation rotation and the iSk state emerges. The external magnetic bias field and disk sample diameter values at which the discontinuities occur are consistent with equilibrium~\cite{Beg2015} and hysteretic behaviour~\cite{Beg2015, Carey2016} studies.
\begin{figure*}
  \includegraphics{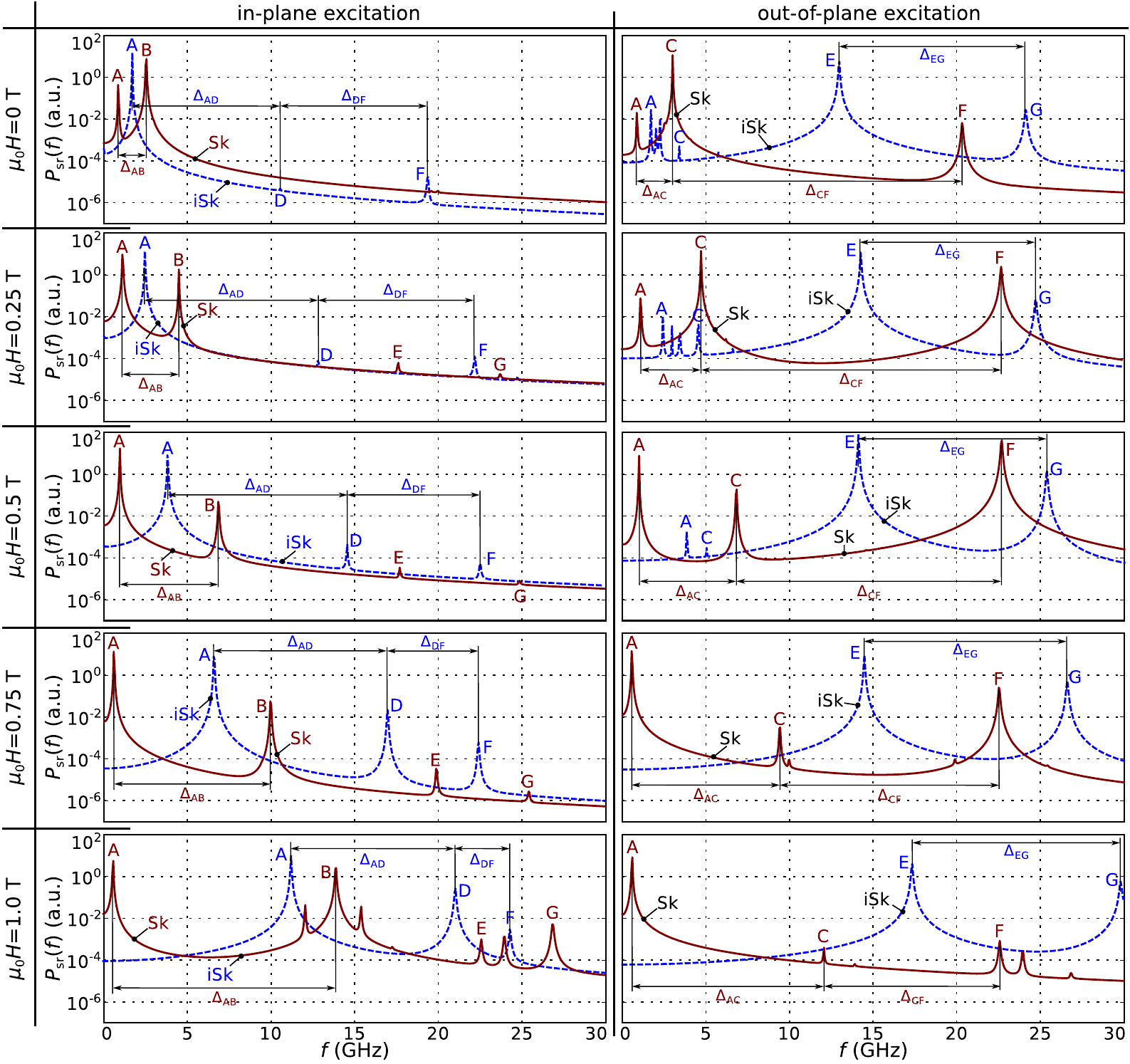}
  \caption{\label{fig:comparison} Comparisons of power spectral densities (PSDs) of ground incomplete skyrmion (iSk) state (solid red line) and metastable isolated skyrmion (Sk) state (dashed blue line) in a $100 \,\text{nm}$ disk sample with $10 \,\text{nm}$ thickness at different values of external magnetic field $H$, computed for an in-plane (left column) and an out-of-plane (right column) excitation.}
\end{figure*}

For an in-plane excitation, in the $H$-dependent PSD map, shown in Fig.~\ref{fig:sk_sweep}(a), five previously discussed eigenmodes (A, B, D, E, and G) are visible in the $H$ range where the Sk$\downarrow$ state is in an equilibrium. The frequency of gyrotropic eigenmode A firstly increases, reaches its maximum at approximately $0.15 \,\text{T}$, and then decreases down to its minimum at approximately $0.9 \,\text{T}$, after which it keeps increasing with $H$. In comparison to the other eigenmodes, its frequency varies over a much smaller range (less than $1 \,\text{GHz}$) over the entire $H$ range where the Sk$\downarrow$ state is in an equilibrium. Similar to the incomplete skyrmion state, the frequency of gyrotropic eigenmode A approaches zero near the switching field $\mu_{0}H_\text{s} = -0.24 \,\text{T}$, suggesting that this eigenmode might govern the isolated skyrmion reversal process. The eigenmode B frequency increases approximately linearly up to $0.6 \,\text{T}$, after which it continues increasing nonlinearly. The frequency of eigenmode D, firstly decreases, reaches its minimum at approximately $0.22 \,\text{T}$, and then continues increasing nonlinearly with $H$. The frequencies of eigenmodes E and G exhibit more complicated behaviour where two extremes (maximum and minimum) are present in their $H$-dependences. When an out-of-plane excitation is used, we observe five previously discussed eigenmodes (A, C, F, H, and J) in the $H$-dependent PSD map, shown in Fig.~\ref{fig:sk_sweep}(b). The eigenmode A now becomes invisible in the PSD map below $0.2 \,\text{T}$. The breathing eigenmode C frequency increases monotonically over the entire Sk$\downarrow$ field range. The frequency dependences of eigenmodes F, H, and J, exhibit more complicated behaviour having both local maximum and minimum in their $H$-dependences.

In the $d$-dependent PSD map, shown in Fig.~\ref{fig:sk_sweep}(c), obtained when an in-plane excitation is used, five previously discussed eigenmodes (A, B, D, E, and G) are present. In contrast to the frequencies of eigenmodes D, E, and G that monotonically decrease with $d$ over a wide range of frequencies, the eigenmodes A and B frequencies vary in a much smaller (less than $1 \,\text{GHz}$) range over entire studied $d$ range. Eigenmodes D, E, and G become invisible in the PSD map below approximately $120 \,\text{nm}$. In Fig.~\ref{fig:sk_sweep}(d), we show the $d$-dependent PSD map for an out-of-plane excitation, where five eigenmodes (A, C, F, H, and J) are visible. Similar to the eigenmodes A and B, the lowest frequency breathing eigenmode frequency changes over a much smaller range than the frequencies of eigenmodes F, H, and J, when the disk sample diameter is changed.

\subsection{Comparison of incomplete skyrmion and isolated skyrmion power spectral densities}
One of the challenges in the study of skyrmionic states in confined helimagnetic nanostructures is the detection of what state emerged in the studied sample. In this subsection we discuss how measuring resonance frequencies can contribute to the identification of the emerged state. Previously, in Sections~\ref{sec:isk} and~\ref{sec:sk}, we studied the dynamics of both incomplete skyrmion (iSk) and isolated skyrmion (Sk) states in disk samples with diameters for which these states were the ground states. Now, we compare the power spectral densities (PSDs) of iSk and Sk states in a $100 \,\text{nm}$ diameter disk sample with $10 \,\text{nm}$ thickness at different external magnetic field values $\mu_{0}H$ (between $0 \,\text{T}$ and $1 \,\text{T}$ in steps of $0.25 \,\text{T}$). In this sample size and at all simulated external magnetic field values, both iSk and Sk states are in equilibrium. More specifically, the Sk state is metastable and the iSk state is the ground state.~\cite{Beg2015} We show the comparison of spatially resolved iSk and Sk PSDs at different external magnetic field values for an in-plane and an out-of-plane excitation in Fig.~\ref{fig:comparison}. Because in a $100 \,\text{nm}$ diameter disk sample there are no dominant iSk and Sk eigenmodes that can be easily detected in experiments above $30 \,\text{GHz}$, we now limit our discussion of PSDs below $30 \,\text{GHz}$ in order to better resolve the key differences, that can contribute to the identification of the present state.

Firstly, in the case of an in-plane excitation (left column in Fig.~\ref{fig:comparison}), the frequency of iSk gyrotropic eigenmode A (the lowest frequency iSk eigenmode), increases with $H$. On the contrary, the Sk gyrotropic eigenmode A (again the lowest frequency Sk eigenmode) frequency remains approximately the same. Furthermore, by increasing the external magnetic field the Sk eigenmode B frequency increases, and consequently, the frequency difference between two lowest frequency Sk eigenmodes $\Delta_\text{AB}$ increases in a wide range of frequencies. In contrast, the frequencies of two lowest frequency iSk eigenmodes A and D both increase with $H$, so that the frequency difference $\Delta_\text{AD}$ between them changes over a small range of frequencies (remains approximately the same). However, at low values of external magnetic field, it could be difficult to measure the iSk eigenmode D due to its relatively small amplitude. In that case, between $0.25 \,\text{T}$ and $0.75 \,\text{T}$, the frequency of dominant iSk eigenmode F does not change, so the $\Delta_\text{AF} = \Delta_\text{AD} + \Delta_\text{DF}$ difference reduces with $H$ for about $5 \,\text{GHz}$.

When we excite the system using an out-of-plane excitation (right column in Fig.~\ref{fig:comparison}), at $H=0$, several resonance frequencies below $5 \,\text{GHz}$ are present, which does not allow a clear identification of the emerged state by measuring resonance frequencies in that region. However, by increasing the external magnetic field, the low frequency part of PSDs simplifies. More specifically, the Sk eigenmode A frequency again does not change, while the Sk breathing eigenmode C, and therefore the difference $\Delta_\text{AC}$, increase with $H$. In addition, for a Sk state above $0.25 \,\text{T}$, the frequency of eigenmode F remains approximately the same, and therefore, the difference $\Delta_{CF}$ decreases with $H$. On the contrary, iSk eigenmodes A, B, and C disappear from the PSD after $\mu_{0}H = 0.5 \,\text{T}$, whereas the frequency difference $\Delta_\text{EG}$ between two most dominant iSk eigenmodes E and G remains approximately the same, since their frequencies both increase.

The dependences of resonant frequencies in this sample with $d=100 \,\text{nm}$ are in a good agreement with the PSD maps shown in Fig.~\ref{fig:isk_sweep}(a) and Fig.~\ref{fig:sk_sweep}(a) and eigenvalue computed results in Supplementary Section S2. This suggests that these identification differences can probably be applied to different sample sizes. At $\mu_{0}H=1 \,\text{T}$, we approach the transition from Sk to iSk state and additional peaks in Sk state PSDs, shown in Fig.~\ref{fig:comparison}, occur.

\subsection{Simulations with real FeGe damping}
In the previous analysis of skyrmionic states dynamics, we intentionally used the small Gilbert damping value $\alpha'=0.002$ as used in other eigenmode studies,~\cite{Kim2014} in order to allow enough separation between peaks in the power spectral densities (enabled by the reduced linewidth) and identify all eigenmodes that can be excited using a particular experimentally feasible excitation. However, in experiments, which eigenmodes can be observed strongly depends on the real value of Gilbert damping. Therefore, in this section, we measure the FeGe Gilbert damping value $\alpha$ and repeat our simulations in order to determine what eigenmodes are expected to be experimentally observed in helimagnetic FeGe confined nanostructures.

\begin{figure}
  \includegraphics{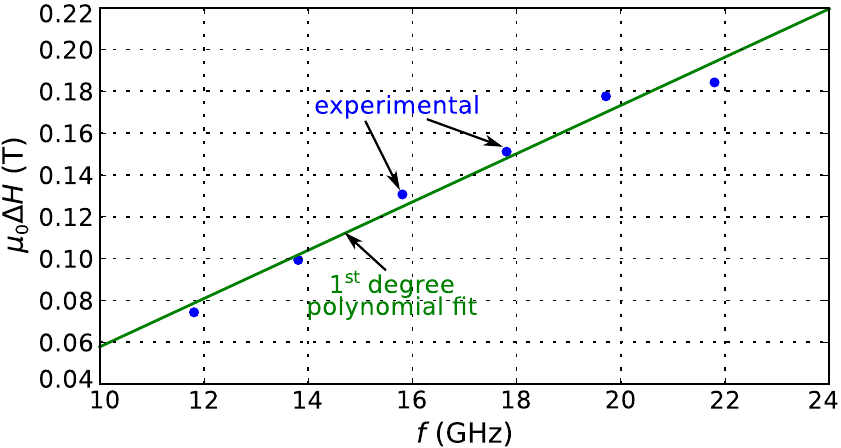}
  \caption{\label{fig:fege_damping_measurements} The linewidth $\Delta H$ (half width at half maximum) measurement points at different resonance frequencies $f$ for a FeGe thin film, together with a first degree polynomial fit from which the Gilbert damping was extracted.}
\end{figure}

We perform the ferromagnetic resonance measurements in a FeGe thin film with $67.8 \pm 0.1 \,\text{nm}$ thickness, grown on the Si substrate in the (111) direction and capped with a $4.77 \pm 0.07 \,\text{nm}$ thin Ge layer.~\cite{Porter2015} We show the linewidth $\Delta H$ (half width at half maximum) measurement points at different resonance frequencies $f$, together with a first degree polynomial fit in Fig.~\ref{fig:fege_damping_measurements}. The polynomial fit allows us to decompose the $\Delta H$ dependence into a frequency independent inhomogeneously-broadened component $\Delta H_{0}$ and an intrinsic damping-related part:~\cite{Liu2003, Kalarickal2006, Kawai2014}
\begin{equation}
  \Delta H = \Delta H_{0} + \frac{\alpha f}{\gamma},
\end{equation}
where $\alpha$ is the Gilbert damping and $\gamma$ is the gyromagnetic ratio. From the slope of the polynomial fit and using the frequency-dependent term that reflects the ``viscous'' damping of the precessive magnetisation motion associated with the FMR, we find $\alpha=0.28 \pm 0.02$.
\begin{figure*}
  \includegraphics{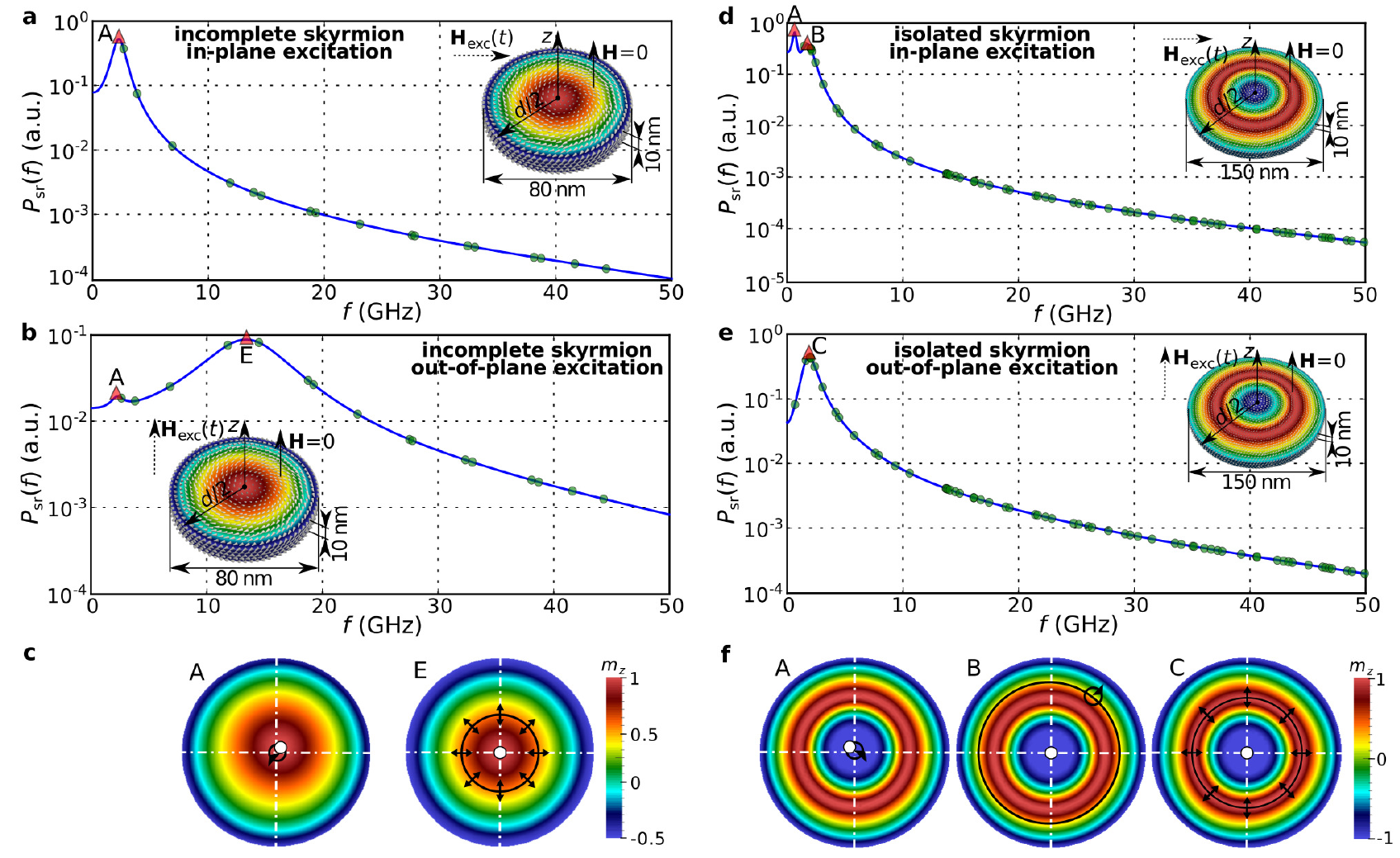}
  \caption{\label{fig:fege_damping_psd} The spatially resolved power spectral densities (PSDs) of an incomplete skyrmion state in a $80 \,\text{nm}$ diameter FeGe disk sample with $10 \,\text{nm}$ thickness at zero external magnetic bias field for (a)~in-plane and (b)~out-of-plane excitation direction. The isolated skyrmion state in a $150 \,\text{nm}$ diameter thin film disk with $10 \,\text{nm}$ thickness at $H=0$ when the system is excited with (c)~in-plane and (d)~out-of-plane excitation. The PSDs are computed using the experimentally measured value of FeGe Gilbert damping $\alpha=0.28$.}
\end{figure*}
\begin{figure}
  \includegraphics{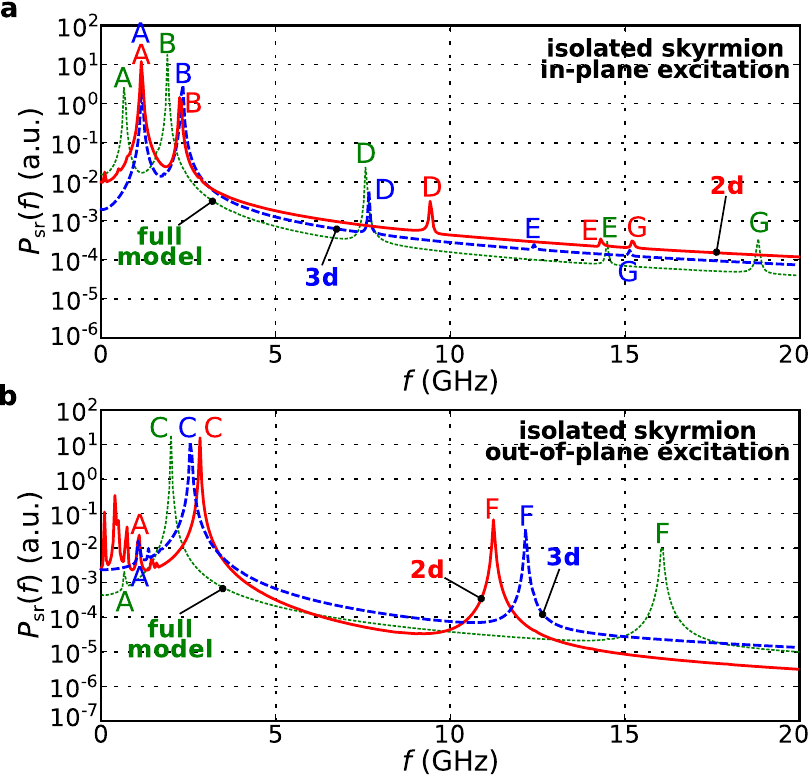}
  \caption{\label{fig:sk_nd2d_psd} The comparison of power spectral densities computed using three-dimensional and two-dimensional models in absence of demagnetisation energy contribution with the PSD obtained using a full simulation model for an isolated skyrmion state in the case of (a)~in-plane and (b)~out-of-plane excitation. Simulated sample is a $150 \,\text{nm}$ diameter disk with $10 \,\text{nm}$ thickness at zero external magnetic field.}
\end{figure}

Now, we use the measured $\alpha=0.28$ and repeat the ringdown simulations for the two skyrmionic ground states that can exist in the studied system. We show the spatially resolved power spectral density of an incomplete skyrmion state in a $80 \,\text{nm}$ diameter disk sample at zero external magnetic bias field for an in-plane and an out-of-plane excitation in Figs.~\ref{fig:fege_damping_psd}(a) and \ref{fig:fege_damping_psd}(b), respectively. We observe that, when the system is excited using an in-plane excitation, only the gyrotropic eigenmode A is present in the PSD. On the other hand, for an out-of-plane excitation, we identify two eigenmodes in the PSD shown in Fig.~\ref{fig:fege_damping_psd}(b). The first one is the gyrotropic eigenmode A, which is also present in the in-plane PSD, and another one is the lowest frequency breathing eigenmode E. The PSDs of the isolated skyrmion state in a $150 \,\text{nm}$ diameter disk sample at zero external magnetic bias field are shown in Figs.~\ref{fig:fege_damping_psd}(c) and \ref{fig:fege_damping_psd}(d) for an in-plane and an out-of-plane excitation, respectively. Now, only the three lowest frequency isolated skyrmion eigenmodes are present. The gyrotropic eigenmode A and eigenmode B can be identified when the system is excited using an in-plane excitation. On the other hand, for the out-of-plane excitation, only the breathing eigenmode C is present.

\subsection{Demagnetisation energy and out-of-plane magnetisation variation effects}
Usually, in the simulations of skyrmionic states dynamics in helimagnetic samples, for simplicity, the demagnetisation energy contribution is neglected and/or a helimagnetic thin film sample is modelled using a two-dimensional mesh. It has been shown that the demagnetisation energy contribution~\cite{Beg2015} and the magnetisation variation in the out-of-film direction~\cite{Rybakov2013, Beg2015} radically change the energy landscape. Consequently, using these assumptions when the static properties of skyrmionic states are explored is not justified. Because of that, in this section, we investigate how these two assumptions affect the dynamics of the isolated skyrmion (Sk) state in studied helimagnetic nanostructure. Firstly, we repeat the isolated skyrmion state simulations in a $150 \,\text{nm}$ diameter disk sample at zero external magnetic bias field, but this time we set the demagnetisation energy contribution $w_{\text{d}}$ in Eq.~(\ref{eq:total_energy}) artificially to zero. Secondly, again in the absence of demagnetisation energy contribution, we simulate the Sk state dynamics under the same conditions, but this time using a two-dimensional mesh to model a thin film sample (i.e. not allowing the magnetisation variation in the out-of-film direction). We show the comparison of power spectral densities computed using three-dimensional and two-dimensional models in absence of demagnetisation energy contribution with the one computed using a full model in Figs.~\ref{fig:sk_nd2d_psd}(a) and \ref{fig:sk_nd2d_psd}(b), for an in-plane and an out-of-plane excitations, respectively.

We observe that although the magnetisation dynamics of identified eigenmodes do not change significantly, the resonance frequencies of some eigenmodes change substantially. In the 3D simulations in absence of demagnetisation energy, while the frequency of eigenmode D remains approximately the same, the frequencies of eigenmodes A and B increase by 71\% and 18\%, respectively. On the other hand, the frequencies of eigenmodes E and G decrease by 14\% and 21\%, respectively. Furthermore, power spectral densities in  Fig.~\ref{fig:sk_nd2d_psd}(b), computed for the out-of-plane excitation, show that the frequency of breathing eigenmode C increases by 17\%, whereas the frequency of eigenmode F decreases by 34\%.

If the thin film sample is modelled using a 2D mesh, which does not allow the magnetisation to vary in the $z$ direction, the frequencies of lateral eigenmodes A, B, and G do not change significantly in comparison to the 3D model in absence of demagnetisation energy contribution. Although the frequency of eigenmode D does not change in the 3D simplified ($w_\text{d} = 0$) model, neglecting the sample thickness, increases its frequency by 20\%. The frequency of eigenmode E increases so that it is approximately the same as in the full 3D model. In comparison to the 3D simplified model, the frequency of breathing eigenmodes C and F further increase by 19\% and 7\%, respectively. In the low frequency region of Fig.~\ref{fig:sk_nd2d_psd}(b) we observe several eigenmodes that are not present in the three-dimensional model. Although there are theoretically less eigenmodes in a two-dimesional sample (the number of existing eigenmodes equals to the number of degrees of freedom in the system), we do not know at what frequencies they will occur. We believe the reason for this is that although the state is the same, its dynamics quilitatively changes at low frequencies due to the missing thickness dimension. We also observe eigenmodes computed using the eigenvalue method that agree with the peaks in PSD shown in Fig.~\ref{fig:sk_nd2d_psd}(b).

\section{Discussion and conclusion}
Using the eigenvalue method, we computed all eigenmodes with frequencies below $50 \,\text{GHz}$ for the incomplete skyrmion, isolated skyrmion, and target states in helimagnetic thin film disk samples at zero external magnetic field. Because which eigenmodes are present in the power spectral density strongly depends on the excitation used to perturb the system from its equilibrium state, we performed the ringdown simulations using two different experimentally feasible excitations (in-plane and out-of-plane). We demonstrated that in all three simulated states, two lateral and one breathing low-frequency eigenmodes exist, as previously demonstrated in two-dimensional skyrmion lattice simulations~\cite{Mochizuki2012} and microwave absorption measurements in bulk helimagnetic materials~\cite{Onose2012, Okamura2013, Schwarze2015}. However, only one lateral eigenmode is gyrotropic, with the skyrmionic state core gyrating around its equilibrium position. The other lateral eigenmode we observed is not gyrotropic because it consists of a single contour ring (defined by the magnetisation $z$ component distribution) revolving around the static skyrmionic state core. The existence of only one gyrotropic eigenmode is in accordance with the recent analytic (rigid skyrmion two-dimensional model) findings by Guslienko and Gareeva,~\cite{Guslienko2016} but in contrast to magnetic bubbles where two gyrotropic eigenmodes were found.~\cite{Makhfudz2012, Moutafis2009, Buttner2015} Because the two gyrotropic eigenmodes with opposite gyration direction in a magnetic bubble imply it possesses mass, our findings suggest that the confined DMI-induced skyrmionic states in the studied system are massless. The low-frequency breathing eigenmode we observe, where a single magnetisation $z$ component contour ring shrinks and expands periodically, is in accordance with findings in Ref.~\onlinecite{Mochizuki2012, Onose2012, Okamura2013, Kim2014, Schwarze2015, Wang2015a, Zhang2015a}.

For the incomplete skyrmion and the isolated skyrmion states we found that the resonance frequencies depend nonlinearly on both the disk sample diameter and the external magnetic bias field. We observed that the frequency of the gyrotropic eigenmode approaches zero near the switching field $H_\text{s}$ (where the reversal of skyrmionic state core occurs) for both incomplete skyrmion and isolated skyrmion states, suggesting that this eigenmode might be the reversal mode of studied skyrmionic states. We find that when the skyrmionic state core orientation reverses, the revolving direction of all lateral eigenmodes changes, which confirms that the revolving direction depends on the direction of the gyrovector as shown in Ref.~\onlinecite{Mochizuki2012, Guslienko2016}.

After we identified all existing eigenmodes of iSk and Sk ground states, we compared their PSDs in the same sample at different external magnetic field values. We identified several characteristics that can contribute to the experimental identification of the state that emerged in the sample by measuring the resonance frequencies.

In the identification and analysis of eigenmodes, we used a small Gilbert damping value in order to provide enough separation between peaks in the PSD. However, which eigenmodes are expected to be observed in experiments strongly depends on the real Gilbert damping value $\alpha$. Therefore we measured $\alpha$ in the FeGe thin film, and carried out ringdown simulations with this $\alpha$. We showed that for the incomplete skyrmion, two eigenmodes (gyrotropic and breathing) are present in the PSD computed using an out-of-plane excitation, whereas only the gyrotropic eigenmode is present in the PSD computed using an in-plane excitation. In the isolated skyrmion case, two lateral eigenmodes are present in the PSD obtained using an in-plane excitation, whereas a single breathing eigenmode is present in the PSD computed after using the out-of-plane excitation.

Our simulations took into account the demagnetisation energy contribution, which is usually neglected for simplicity in both analytic and simulation works. To explore the importance of model assumptions, we carried out further systematic simulation studies in which we set the demagnetisation energy contribution artificially to zero. We also repeated the simulations under the same conditions on 3D and 2D meshes (with and without permissible magnetisation variation in the out-of-film direction, respectively). We found that although the magnetisation dynamics of eigenmodes does not change significantly, their frequencies change substantially. This suggests that ignoring the demagnetisation energy contribution or approximating a thin film helimagnetic sample using a two-dimensional mesh is not always justified.

This work provides a systematic dynamics study of skyrmionic states in confined helimagnetic nanostructures. We report all eigenmodes present in the sample as well as which eigenmodes can be observed using particular experimentally feasible excitations. Apart from contributing to fundamental physics, this work could support experimentalists to determine what magnetisation configuration is present in the confined helimagnetic sample by measuring ferromagnetic resonance spectra.

All data supporting this study are openly available from the University of Southampton repository at http://doi.org/10.5258/SOTON/403976.

\begin{acknowledgments}
  This work was financially supported by the EPSRC Doctoral Training Centre (DTC) grant EP/G03690X/1, OpenDreamKit – Horizon 2020 European Research Infrastructure project (676541), and the EPSRC Programme grant on Skyrmionics (EP/N032128/1). D.C.-O. acknowledges the financial support from CONICYT Chilean scholarship programme Becas Chile (72140061). C.C. acknowledges the support from a Junior Research Fellowship at Gonville and Caius College. The scholarship of C.S.S was supported by the STFC Rutherford Appleton Laboratory and the Hitachi Cambridge Laboratory. C.H.M. acknowledges support by EPSRC through the standard research grant EP/J007110/1. We also acknowledge the use of the IRIDIS High Performance Computing Facility, and associated support services at the University of Southampton, in the completion of this work. We thank Robert L. Stamps for useful discussions.
\end{acknowledgments}


%

\clearpage


\section*{Supplementary material}

# Supplementary Information: Dynamics of skyrmionic states in confined helimagnetic nanostructures

Marijan Beg,[1, *] Maximilian Albert,[1] Marc-Antonio Bisotti,[1] David Cortés-Ortuño,[1]
Weiwei Wang,[1, 2] Rebecca Carey,[1] Mark Vousden,[1] Ondrej Hovorka,[1] Chiara
Ciccarelli,[3] Charles S. Spencer,[4] Christopher H. Marrows,[4] and Hans Fangohr[1, †]

[1] *Faculty of Engineering and the Environment, University of Southampton, Southampton, SO17 1BJ, United Kingdom*

[2] *Department of Physics, Ningbo University, Ningbo, 315211, China*

[3] *Cavendish Laboratory, University of Cambridge, Cambridge, CB3 0HE, United Kingdom*

[4] *School of Physics and Astronomy, University of Leeds, Leeds, LS2 9JT, United Kingdom*

## I. SUPPLEMENTARY SECTION S1: TARGET (T) STATE

In this Supplementary Section, we study the dynamics of the Target (T) state. This state is characterised by the scalar value $S_a$ being in the $[1, 2]$ range,[1] because the magnetisation component $m_z$ covers the $[-1, 1]$ range twice, plus an additional magnetisation tilting at the boundary due to the specific boundary conditions.[2] Although the T state is never the ground state (global energy minimum) in the studied system, it is in an equilibrium[1] for the disk sample diameters $144\,\text{nm} \leq d \leq 180\,\text{nm}$ and the external magnetic bias field $\mu_0 H \leq 0.24\,\text{T}$. More precisely, for these $d$ and $H$ values, target state is in the local and not global energy minimum state. We explore the target state dynamics in a $160\,\text{nm}$ diameter disk sample with $10\,\text{nm}$ thickness at zero external magnetic bias field. In order to obtain the target state in the sample, the magnetisation is initialised with a state having similar configuration as explained in Ref. 1.

In the first step, we compute all existing eigenmodes below $50\,\text{GHz}$. Using the ringdown method, we compute the Spatially Averaged (SA) and Spatially Resolved (SR) Power Spectral Densities (PSDs) for an in-plane excitation, overlaid with the eigenvalue results, and show them in Supplementary Fig. 1 (a) and Supplementary Fig. 1 (b), respectively. In these power spectral densities, we observe eleven peaks (eigenmodes A, B, C, D, E, F, G, I, J, L, and M) and show their schematic representations in Supplementary Fig. 1 (e). We identify the two lowest frequency peaks at $0.72\,\text{GHz}$ and $0.89\,\text{GHz}$, which correspond to the gyrotropic eigenmode A and lateral eigenmode B, respectively. Similar to the incomplete skyrmion and isolated skyrmion states, the gyrotropic eigenmode A magnetisation dynamics consists of a target state core (where $m_z = -1$) gyrating around its equilibrium position in the CounterClockWise (CCW) direction. The magnetisation dynamics of eigenmode B is composed of a magnetisation $z$ component contour ring revolving in the ClockWise (CW) direction around the sample centre. At $2.86\,\text{GHz}$, we identify another dominant eigenmode D, which consists of a T state core and a contour ring rotating in the CCW direction, but mutually out-of-phase. In the case of the eigenmode F at $8.17\,\text{GHz}$, the T state core is static, but two magnetisation contour rings revolve mutually out-of-phase in the CW direction. The magnetisation dynamics of all other eigenmodes (G, I, J, L, and M) present in the in-plane power spectral densities have significantly lower power. They are all lateral and consist of different combinations of revolving contour rings and the T state core, and we show their schematic representations in Supplementary Fig. 1 (e). All discussed eigenmodes so far are lateral (in-plane), but in the discussed in-plane PSDs, two breathing eigenmodes C and E are also present with significantly lower power and we will discuss them subsequently when an out-of-plane excitation is used to perturb the system. We show the magnetisation dynamics of all eigenmodes computed using the eigenvalue method and in Supplementary Videos 4 and 5, and their schematic representations in Supplementary Section S3.

Now, we excite the system with an out-of-plane excitation, compute the SA and SR PSDs, and show them in Supplementary Fig. 1 (c) and Supplementary Fig. 1 (d), respectively. In this case, we observe five dominant breathing eigenmodes (C, E, H, K, and N) and significantly weaker lateral eigenmodes (A, B, D, F, G, I, J, and L) discussed previously. The lowest frequency breathing eigenmode C at $1.60\,\text{GHz}$ contains a single magnetisation $z$ component contour ring that shrinks and expands periodically. Another two dominant peaks in these PSDs correspond to eigenmodes E and H at $5.59\,\text{GHz}$ and $18.93\,\text{GHz}$, respectively. The eigenmode E magnetisation dynamics consists of two contour rings breathing mutually out-of-phase. Similarly, the magnetisation dynamics of the eigenmode H consists of three breathing contour rings where the middle contour ring breathes out-of-phase with respect to the inner and the outer one. The eigenmodes K and N magnetisation dynamics consist of different combinations of breathing contours and we show their schematic representations in Supplementary Fig. 1 (e).

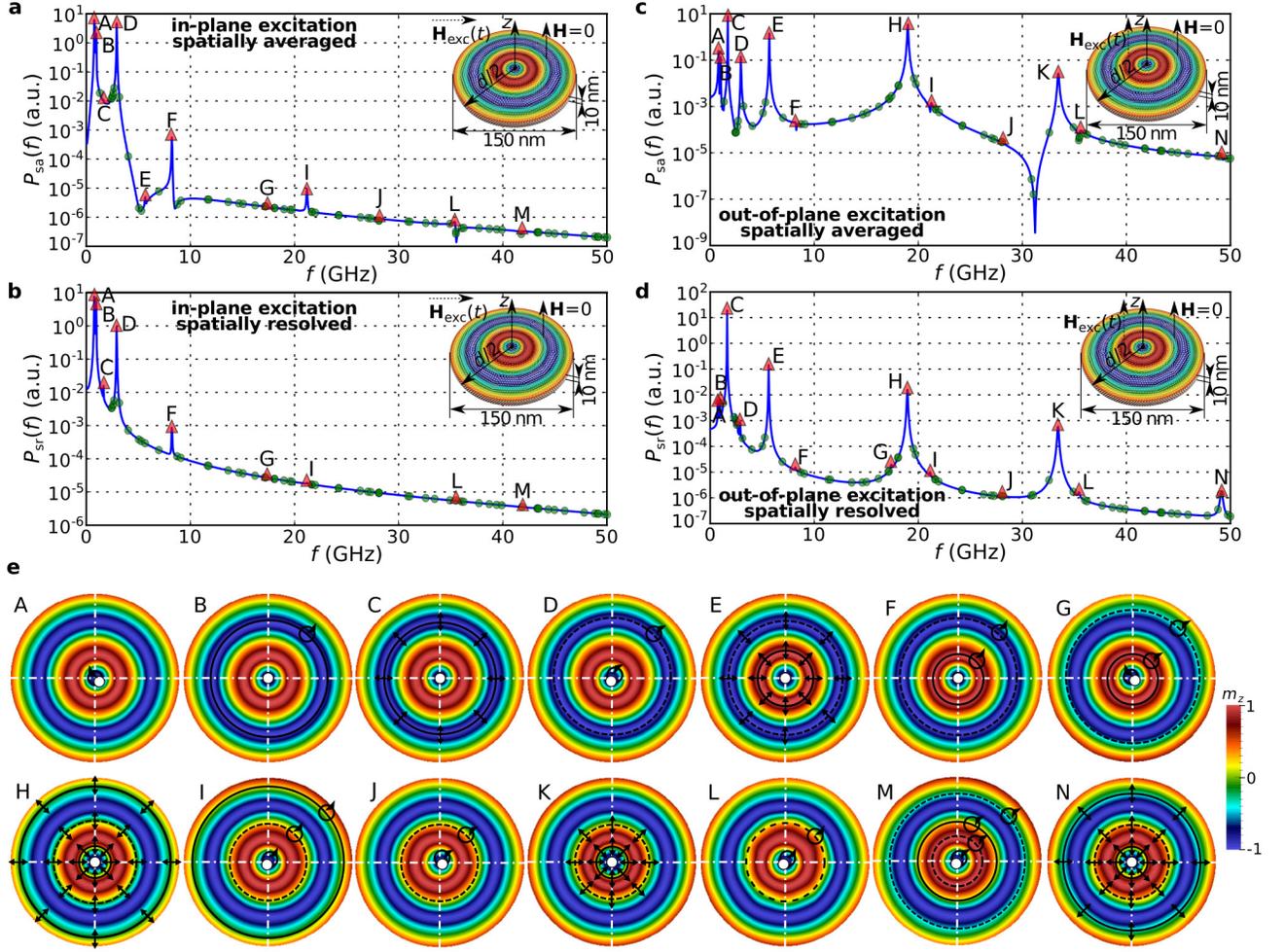

Supplementary Fig. 1. The Power Spectral Densities (PSDs) of a Target (T) equilibrium state in a 160 nm diameter FeGe disk sample with 10 nm thickness at zero external magnetic bias field. (a) Spatially averaged and (b) spatially resolved PSDs for an in-plane excitation, together with overlaid resonance frequencies computed using the eigenvalue method. The resonant frequencies obtained using the eigenvalue method are marked with a triangle symbol (△) if they can be activated using a particular excitation and with a circle symbol (◦) otherwise. (c) Spatially averaged and (d) spatially resolved PSDs computed when the target state is perturbed from its equilibrium with an out-of-plane excitation. (e) Schematic representations of magnetisation dynamics associated with the identified eigenmodes. Schematically, we represent the target state core with a circle symbol, together with a directed loop if it gyrates around its equilibrium position. Contour rings represented using dashed lines revolve/breathe out-of-phase with respect to the those marked with solid lines. The magnetisation dynamics animations of all identified eigenmodes are provided in Supplementary Videos 4 and 5.

## II. SUPPLEMENTARY SECTION S2: EXTERNAL MAGNETIC FIELD AND DISK SAMPLE DIAMETER DEPENDENCE OF RESONANCE FREQUENCIES COMPUTED USING EIGENVALUE METHOD

For every ringdown simulation in our study, we also carried out the eigenvalue computation. In the Power Spectral Density (PSD) maps we discussed in the main text, some of the eigenmodes become invisible at certain values of external magnetic field $H$ and disk sample diameter $d$ due to their small amplitude. In this Supplementary Section, we show the $H$ and $d$ dependences computed using the eigenvalue method of all identified eigenmodes over the entire range of studied $H$ and $d$ values. We show the dependences of incomplete Skyrmion (iSk) resonance frequencies on $H$ and $d$ in Supplementary Fig. 2 (a) and Supplementary Fig. 2 (b), respectively. Similarly, we show the dependences of isolated Skyrmion (Sk) resonance frequencies on $H$ and $d$ in Supplementary Fig. 3 (a) and Supplementary Fig. 3 (b), respectively. We represent discrete eigenvalue computed resonance frequencies using dots and emphasise the dependences of eigenmodes identified in the main text using a solid line. We find a perfect

agreement between eigenvalue and ringdown methods in $d$ and $H$ ranges where eigenmodes are visible in PSD maps.

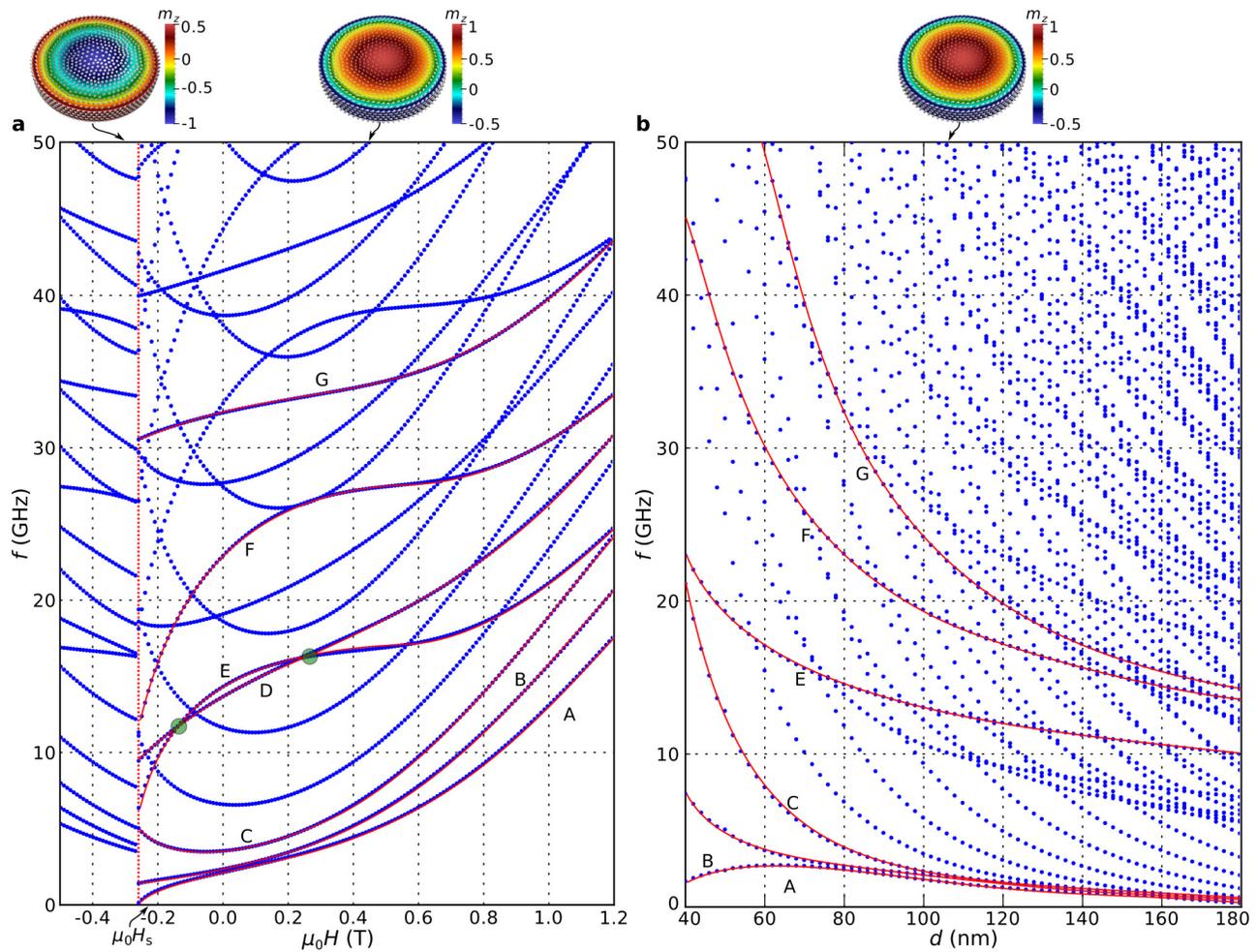

Supplementary Fig. 2. The external magnetic field $H$ and disk sample diameter $d$ dependence of incomplete Skyrmion (iSk) state resonance frequencies computed using the eigenvalue method. Discrete eigenvalue computed resonance frequencies are represented using dots and the dependences of eigenmodes identified in the main text are emphasised using a solid line.

Since we are now able to look at the resonance frequency dependences over the entire range of studied $d$ and $H$ values, we can determine whether any crossings of eigenmodes occur. We observe many crossings between all existing eigenmodes, but we focus only on the eigenmodes that can be excited using an experimentally feasible excitation. In the iSk state, we identify two crossings between eigenmodes E and D in the plot shown in Supplementary Fig. 2 (a) at approximately $-0.13\,\mathrm{T}$ and $0.27\,\mathrm{T}$ with a green circle. On the other hand, there are no crossings in $d$ dependence plot shown in Supplementary Fig. 2 (b). In the Sk dynamics, we find three crossings in the plot shown in Fig. 3 (a). The first one is between eigenmodes E and F at approximately $0.38\,\mathrm{T}$ and the other two are between eigenmodes E and F with eigenmode D near the switching field $\mu_0 H_\mathrm{s}$. Similarly, the eigenmode D crosses eigenmodes E and F at approximately $105\,\mathrm{nm}$ and $100\,\mathrm{nm}$, respectively, in the $d$ dependence plot shown in Fig. 3 (b).

## III. SUPPLEMENTARY SECTION S3: ALL EIGENMODES OF SKYRMIONIC STATES COMPUTED USING THE EIGENVALUE METHOD

In this Supplementary Section, we show the schematic representations of all identified eigenmodes with frequencies below $50\,\mathrm{GHz}$ computed using the eigenvalue method. We show the incomplete Skyrmion (iSk) eigenmodes in Supplementary Fig. 4, isolated Skyrmion (Sk) eigenmodes in Supplementary Figures 5 and 6, and Target (T) eigenmodes in Supplementary Figures 7 and 8.

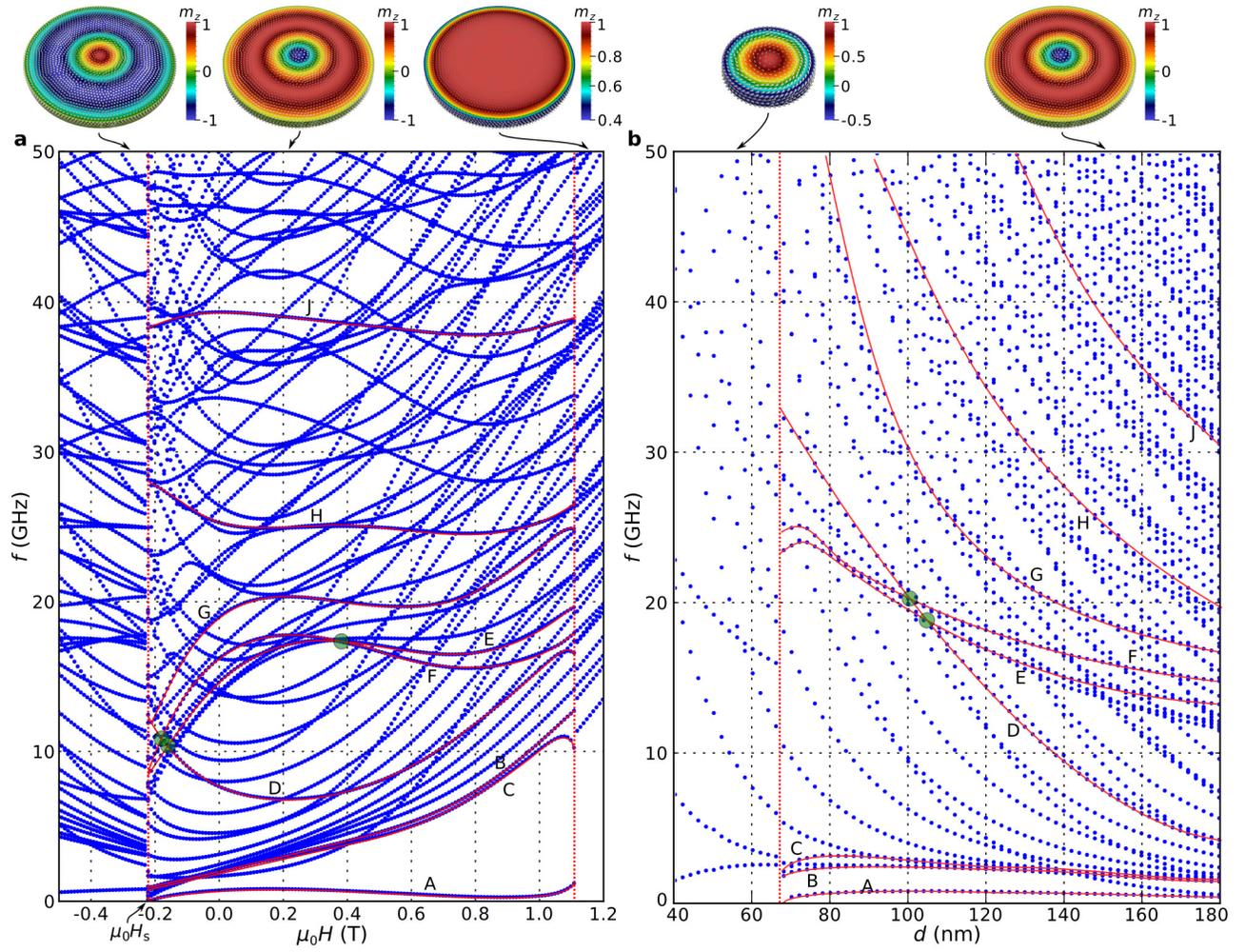

Supplementary Fig. 3. The external magnetic field $H$ and disk sample diameter $d$ dependence of isolated Skyrmion (Sk) state resonance frequencies computed using the eigenvalue method. Discrete eigenvalue computed resonance frequencies are represented using dots and the dependences of eigenmodes identified in the main text are emphasised using a solid line.

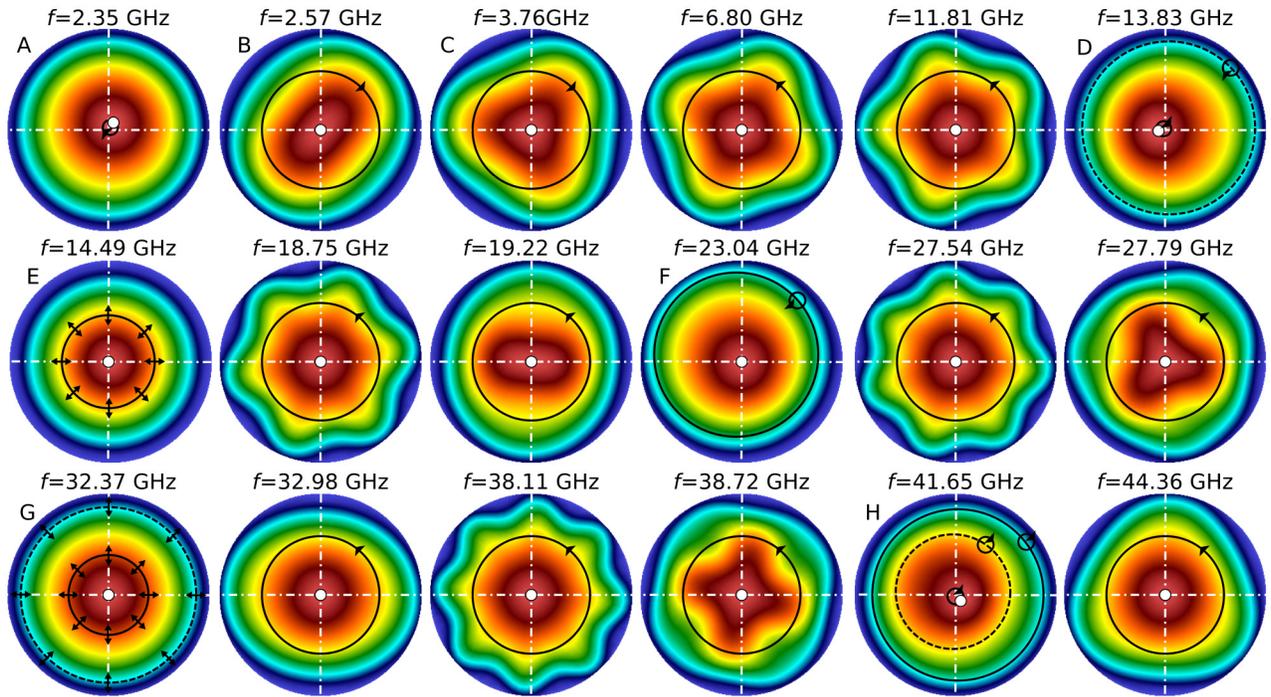

Supplementary Fig. 4. The schematic representation of magnetisation dynamics corresponding to the identified eigenmodes in the $0 - 50$ GHz frequency range for an incomplete Skyrmion (iSk) state, obtained using the eigenvalue method. The simulated sample is a $80$ nm diameter disk sample with $10$ nm thickness at zero external magnetic bias field.

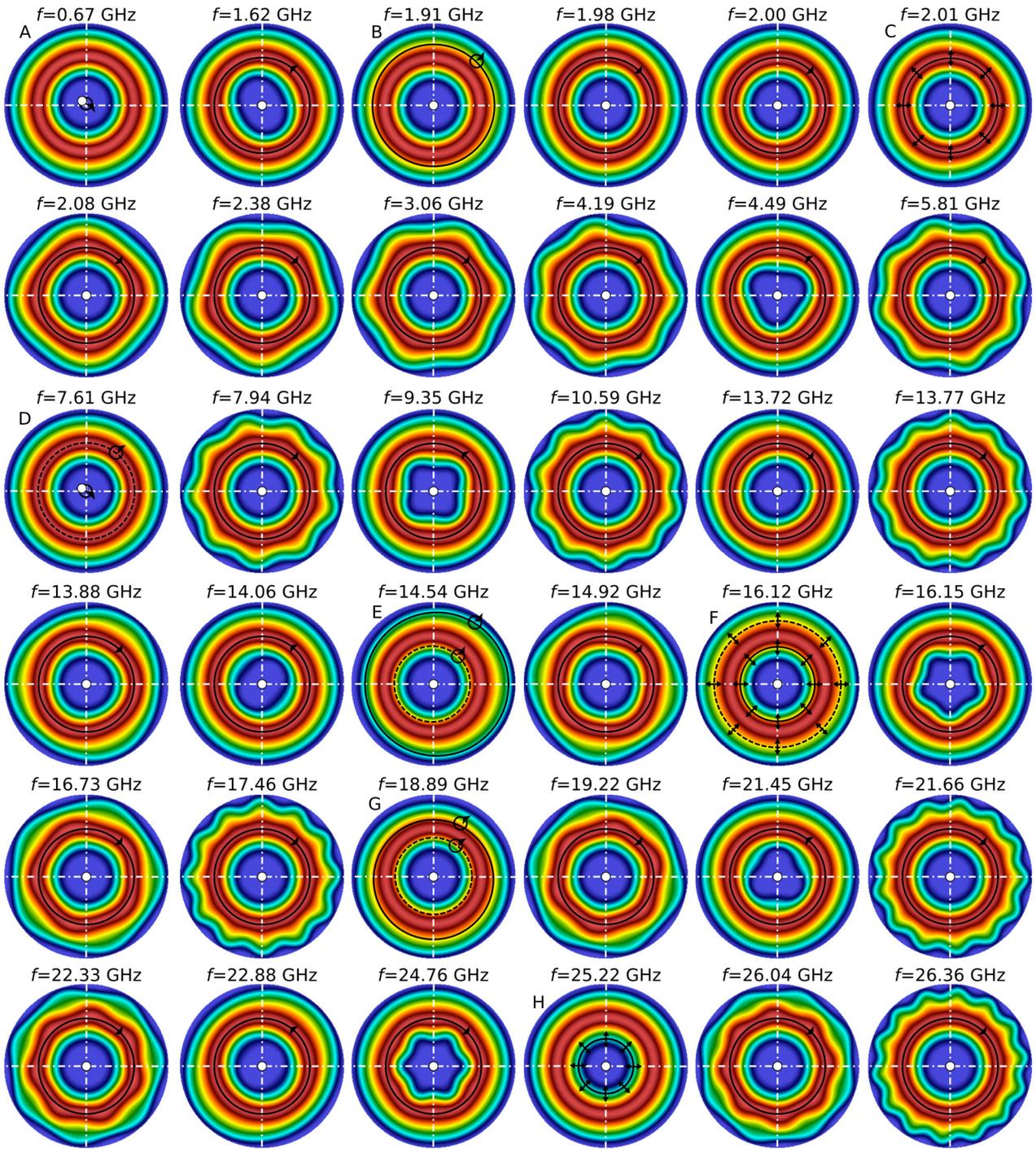

Supplementary Fig. 5. The schematic representation of magnetisation dynamics corresponding to the identified eigenmodes in the 0 − 27 GHz frequency range for an isolated Skyrmion (Sk) state, obtained using the eigenvalue method. The simulated sample is a 150 nm diameter disk sample with 10 nm thickness at zero external magnetic bias field.

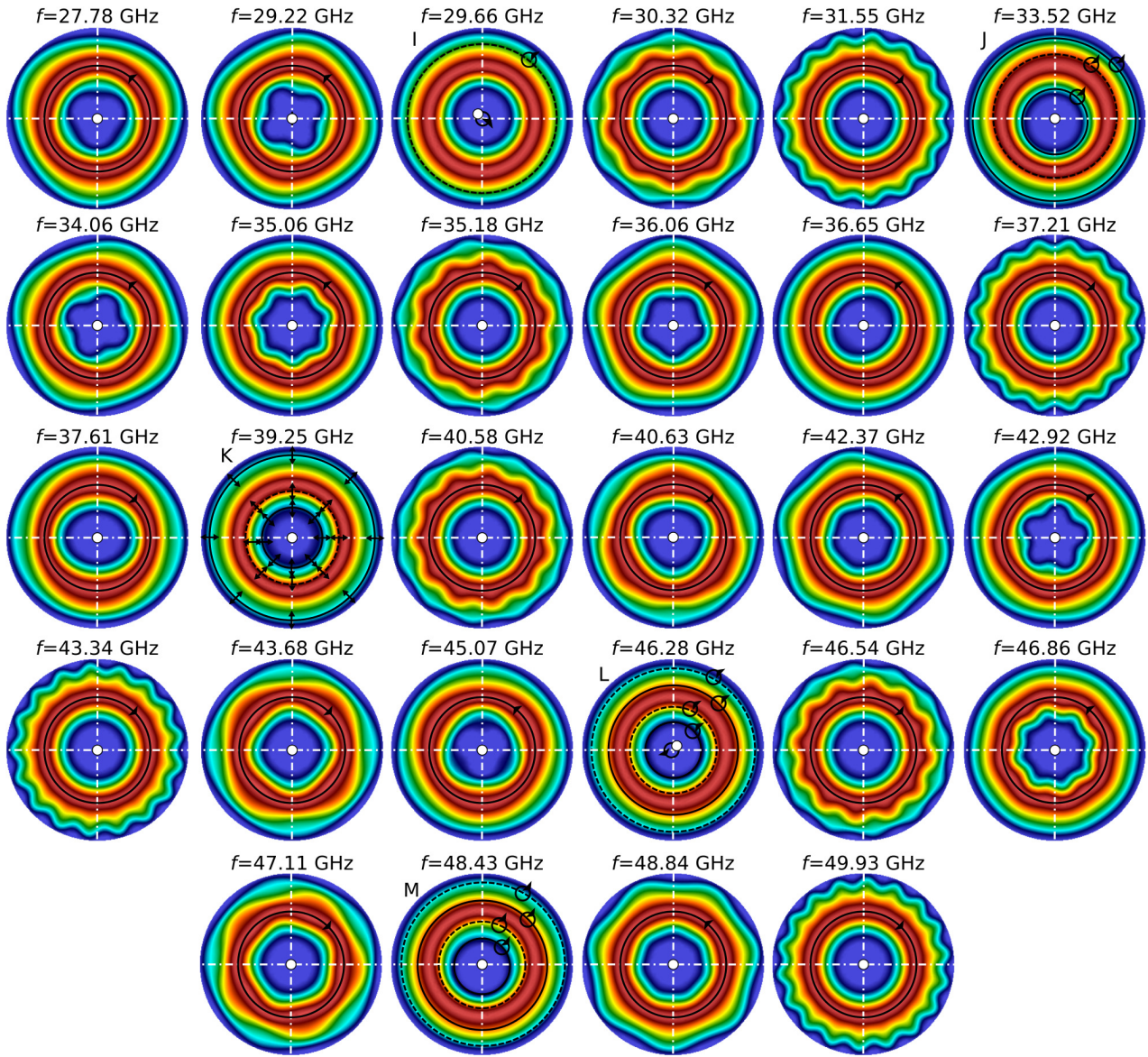

Supplementary Fig. 6. The schematic representation of magnetisation dynamics corresponding to the identified eigenmodes in the $27 - 50$ GHz frequency range for an isolated Skyrmion (Sk) state, obtained using the eigenvalue method. The simulated sample is a 150 nm diameter disk sample with 10 nm thickness at zero external magnetic bias field.

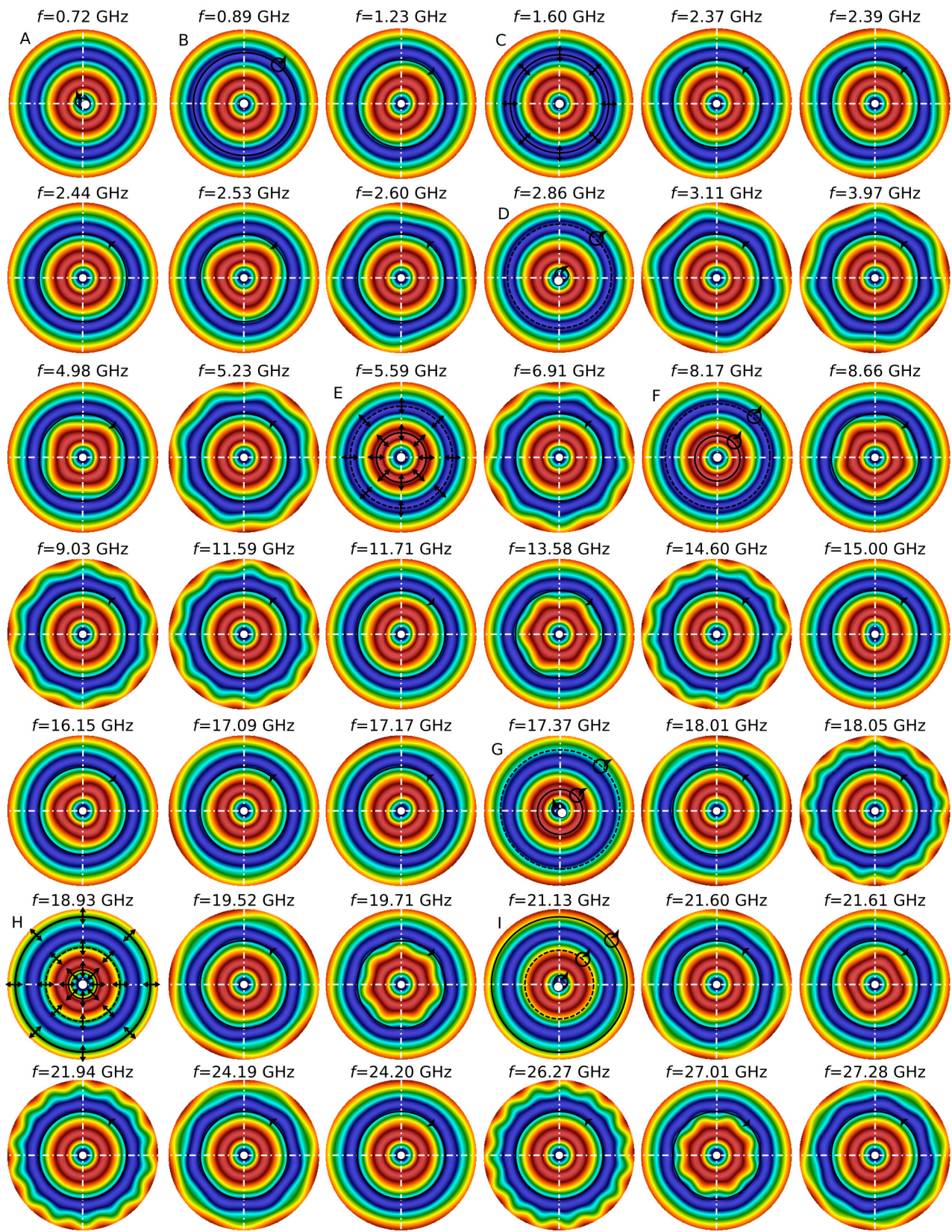

Supplementary Fig. 7. The schematic representation of magnetisation dynamics corresponding to the identified eigenmodes in the $0 - 28$ GHz frequency range for a Target (T) state, obtained using the eigenvalue method. The simulated sample is a $160$ nm diameter disk sample with $10$ nm thickness at zero external magnetic bias field.

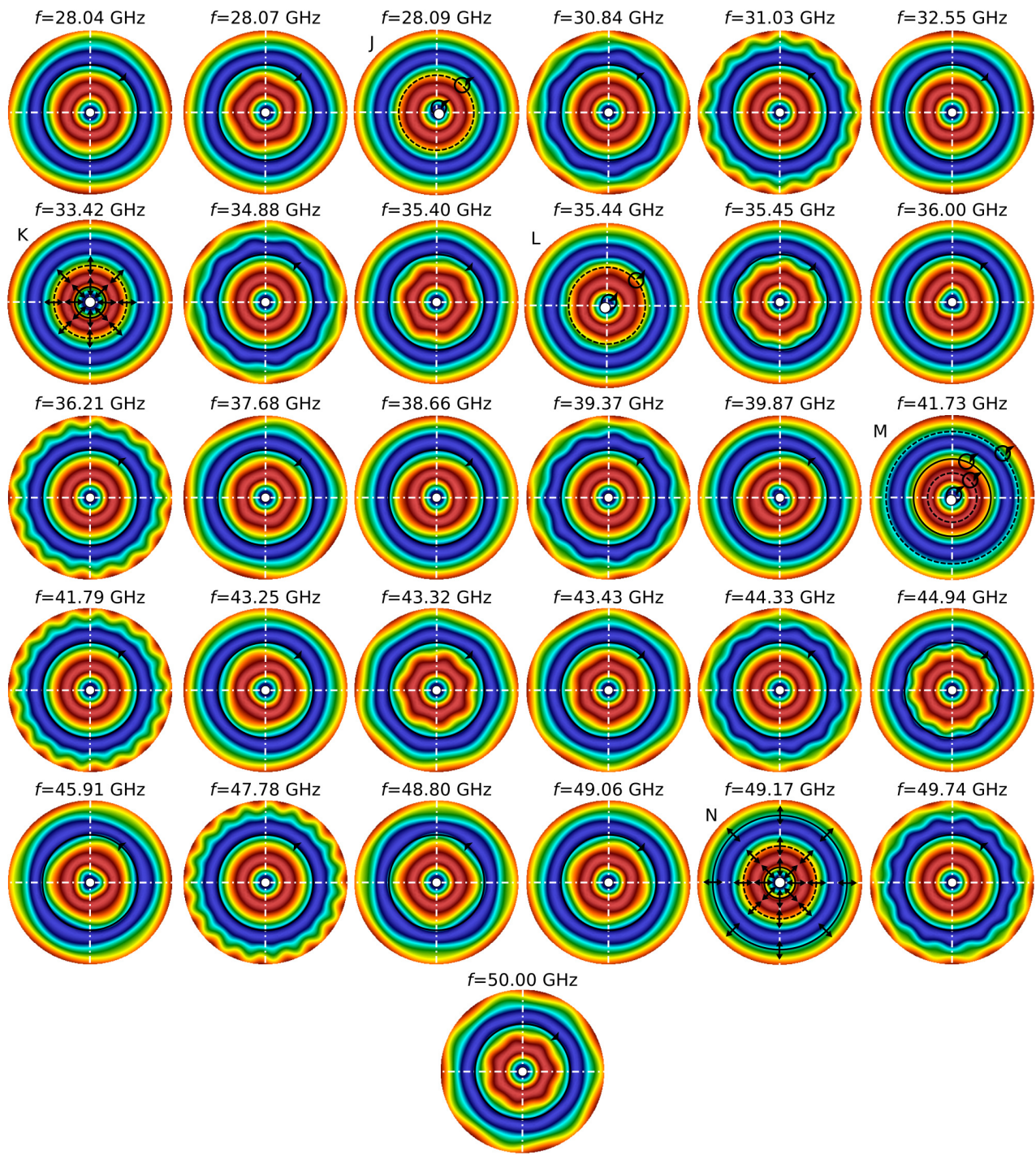

Supplementary Fig. 8. The schematic representation of magnetisation dynamics corresponding to the identified eigenmodes in the $28 - 50\,\mathrm{GHz}$ frequency range for a Target (T) state, obtained using the eigenvalue method. The simulated sample is a $160\,\mathrm{nm}$ diameter disk sample with $10\,\mathrm{nm}$ thickness at zero external magnetic bias field.



\end{document}